\documentclass[manuscript, nonacm]{acmart}
\usepackage[flushleft]{threeparttable}
\setcopyright{none}
\renewcommand\footnotetextcopyrightpermission[1]{}

\usepackage{pifont}   
\usepackage{tikz}
\usepackage{colortbl}
\definecolor{offloadbg}{HTML}{FFF5CC}
\newcolumntype{Y}{>{\columncolor{offloadbg}}c}
\definecolor{dotAssistant}{HTML}{E69F00}
\definecolor{dotFeedback}{HTML}{56B4E9}
\definecolor{dotReward}{HTML}{009E73}
\definecolor{dotEmpty}{HTML}{C9CCD1}
\newcommand{\condDot}[2]{\tikz[baseline=-0.55ex]{\ifnum#2=1 \fill[#1] (0,0) circle (0.55ex);\else \fill[dotEmpty] (0,0) circle (0.55ex);\fi}}
\DeclareRobustCommand{\factorAssistant}{\texorpdfstring{\condDot{dotAssistant}{1}\,}{}}
\DeclareRobustCommand{\factorFeedback}{\texorpdfstring{\condDot{dotFeedback}{1}\,}{}}
\DeclareRobustCommand{\factorReward}{\texorpdfstring{\condDot{dotReward}{1}\,}{}}
\DeclareRobustCommand{\condDots}[3]{\condDot{dotAssistant}{#1}\hspace{0.06em}\condDot{dotFeedback}{#2}\hspace{0.06em}\condDot{dotReward}{#3}\hspace{0.4em}}

\PassOptionsToPackage{dvipsnames}{xcolor}  
\definecolor{ForestGreen}{RGB}{34,139,34}
\definecolor{BrickRed}{RGB}{178,34,34}

\usepackage{tabularx}
\usepackage{framed}
\definecolor{promptbg}{gray}{0.94}
\definecolor{promptink}{gray}{0.42}
\usepackage[most]{tcolorbox}
\newtcolorbox{promptbox}{breakable, enhanced, colback=promptbg, colframe=promptbg,
  boxrule=0pt, arc=0pt, left=6pt, right=6pt, top=4pt, bottom=4pt,
  before skip=6pt, after skip=6pt, fontupper=\small\itshape\color{promptink}}
\newenvironment{excerptbox}{\colorlet{shadecolor}{promptbg}\begin{snugshade*}\small}{\end{snugshade*}}
\newcommand{\excerptcaption}[1]{\par\addvspace{\medskipamount}\noindent\textit{#1}\par\nobreak\smallskip}

\begin{document}

\author{Sebastian Maier}
\email{maier.sebastian@campus.lmu.de}
\authornote{Both authors contributed equally to this research.}
\affiliation{
  \institution{LMU Munich \& Munich Center for Machine Learning (MCML)}
  \city{Munich}
  \country{Germany}
}

\author{Kai Schwabe}
\authornotemark[1] 
\affiliation{
  \institution{LMU Munich}
  \city{Munich}
  \country{Germany}
}

\author{Manuel Schneider}
\affiliation{
  \institution{LMU Munich}
  \city{Munich}
  \country{Germany}
}

\author{Stefan Feuerriegel}
\affiliation{
  \institution{LMU Munich \& Munich Center for Machine Learning (MCML)}
  \city{Munich}
  \country{Germany}
}

\title{Designing Against Deskilling: Metacognitive Feedback Reduces Cognitive Offloading to LLM Assistants}

\begin{abstract}
  Cognitive offloading to AI can reduce opportunities to practice skills, creating risks of deskilling. However, it remains unclear how to prevent deskilling without restricting access to AI. Here, we design two interventions to reduce offloading decisions: (1) metacognitive feedback that makes the implications of offloading for users explicit, and (2) an effort-based reward that incentivizes less extensive LLM assistance. We test both in a preregistered online experiment ($N = 704$) with a 2$\times$2 design and a no-AI control. The task was to practice fraction arithmetic with an LLM-based assistant that provided solutions only on explicit request, followed by an unaided test. Metacognitive feedback reduced answer offloading (OR $= 0.47$) and improved test performance (OR $= 1.51$). We found no evidence that the reward affected either outcome. Our results identify metacognitive feedback as a promising design choice to reduce cognitive offloading.
\end{abstract}



\begin{teaserfigure}
  \centering
  \includegraphics[width=\textwidth]{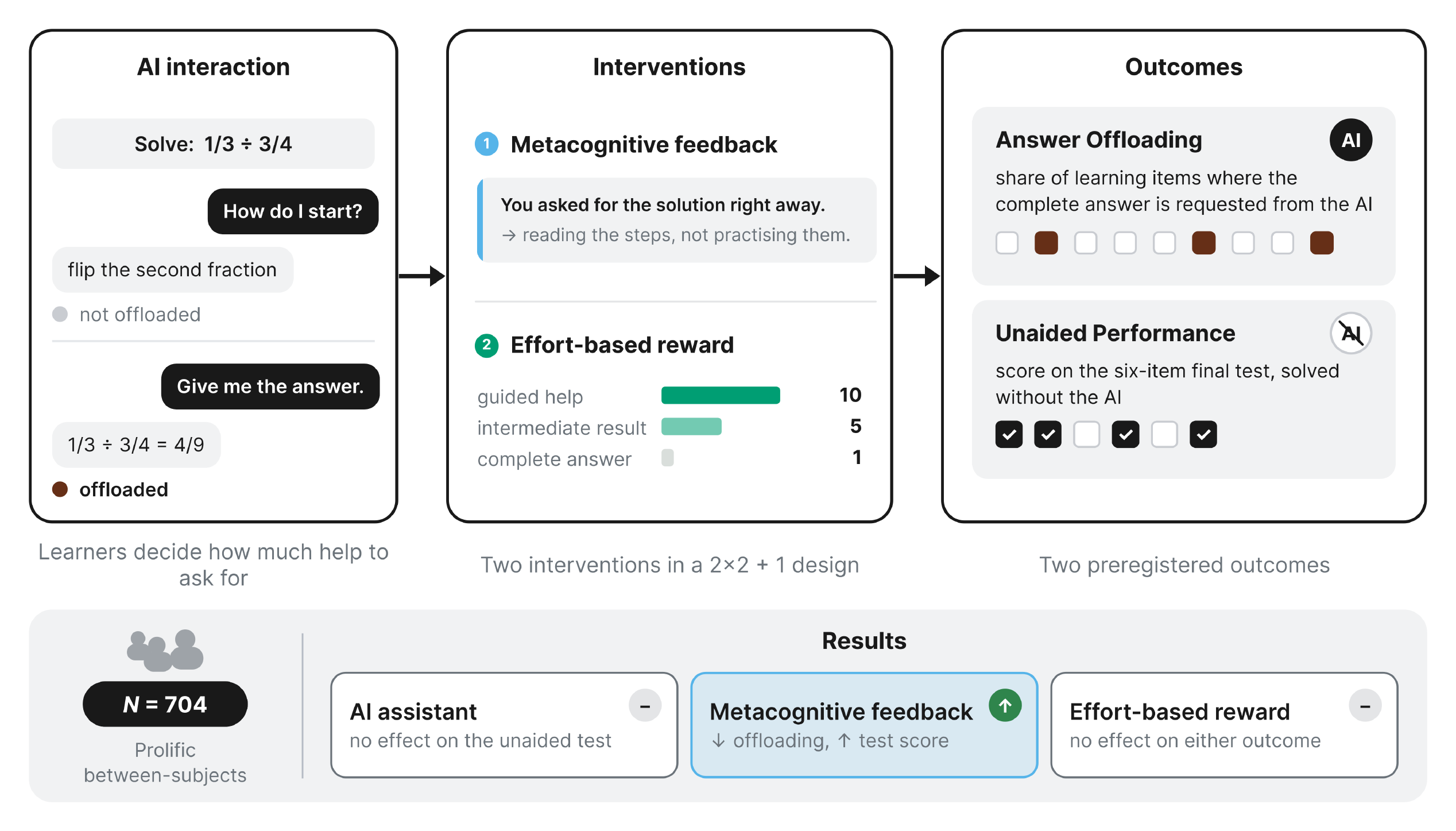}
  \caption{Study overview. Participants are tasked to solve exercises with fraction arithmetic; upon personal choice, the participants may offload the task to an LLM assistant that provides solutions only on request (left). To reduce offloading and subsequent deskilling, we compare two design interventions in a 2$\times$2-plus-control design: (i)~metacognitive feedback and (ii)~an effort-based reward, together  with a no-AI control (center). The preregistered outcomes are answer offloading and unaided test performance as a measure of deskilling (right). Our results show that metacognitive feedback reduced offloading and improved performance; the reward changed neither outcome (bottom).}
  \Description{Three-panel diagram with a results strip below. Left panel, titled AI interaction: an example chat for the item 1/3 divided by 3/4, in which the request ``How do I start?'' receives the hint ``flip the second fraction'' and is marked not offloaded, while ``Give me the answer.'' receives the full result 4/9 and is marked offloaded; the caption beneath reads ``Learners decide how much help to ask for''. Center panel, titled Interventions: intervention 1, metacognitive feedback, with an example screen reading ``You asked for the solution right away.'' followed by an arrow to ``reading the steps, not practicing them.''; intervention 2, the effort-based reward, under which correct answers earn 10 points after guided help, 5 after an intermediate result, and 1 after a complete answer; the caption beneath reads ``Two interventions in a 2 by 2 plus 1 design''. Right panel, titled Outcomes: answer offloading, the share of learning items where the complete answer is requested from the AI, shown as nine squares of which three are marked; and unaided performance, the score on the six-item final test solved without the AI, shown as six squares of which four are checked; the caption beneath reads ``Two preregistered outcomes''. Bottom strip: N = 704 participants, recruited on Prolific, between-subjects; three result cards state that the AI assistant had no effect on the unaided test, metacognitive feedback reduced offloading and increased the test score, and the effort-based reward had no effect on either outcome.}
  \label{fig:teaser}
\end{teaserfigure}

\maketitle

\section{Introduction}

LLM assistants offer a new way to support learning and skill development through natural language interaction. In educational contexts, such LLM assistants can explain unfamiliar concepts, provide feedback on reasoning, and generate individualized problems to practice, for instance by scaffolding programming practice \cite{kazemitabaar2024codeaid} or supporting mathematics learning \cite{liu2026amiwrite}.

However, access to LLM support does not necessarily translate into better skills when the assistance is no longer available. There is growing concern that LLM assistance may even undermine skill development or erode existing skills, commonly discussed as \textit{deskilling} \cite{lenharo2026skills}. For example, high-school students who practiced mathematics with unrestricted ChatGPT scored higher during practice yet lower on the subsequent exam without LLM assistance \cite{bastani2025guardrails}. Similarly, experienced endoscopists detected fewer adenomas when performing colonoscopies without AI after months of AI-assisted practice \cite{budzyn2025deskilling}, and LLM assistance during creative tasks lowered subsequent independent creativity \cite{kumar2025creativity}. One possible explanation is how people use the LLM assistance, particularly whether they work through a problem with support or delegate the problem-solving to the LLM.

How people use a technology can shape whether it supports or harms skill development \cite{salomon1991partners}. A useful lens for understanding these differences is \emph{cognitive offloading}, which involves using actions or external tools to change a task’s processing requirements and reduce cognitive demand \cite{risko2016offloading}. For example, a student can request the complete answer (which we call \emph{answer offloading} throughout the rest of the paper), thereby removing exactly the effortful processing on which durable learning depends \cite{bjork2011difficulties, bastani2025guardrails, gajos2022learning}. The same student could instead ask how to approach the problem when stuck or request a deeper explanation, which are forms of instrumental help that support continued engagement with the task and thus may facilitate learning \cite{nelsonlegall1981help, lehmann2024classroom, kumar2025mathtutor, varone2026motivations}. These different ways of engaging with LLM assistance may have different consequences for learning, yet little is known about how interaction design can influence a user's offloading decisions without restricting the assistance available.

To reduce offloading, existing interventions have mainly changed what output the assistant provides. A common approach is to use guardrails that constrain the assistant toward tutoring behavior by offering explanations and guiding questions while withholding direct solutions. Such approaches can preserve later performance without LLM support \cite{bastani2025guardrails, bassner2025dissociation, kazemitabaar2023codegen}. However, restricting assistance also has drawbacks: withholding answers can demotivate lower-skilled learners and increase their dependence on the system \cite{myung2026scaffolding}, while users may circumvent guardrails when unrestricted assistance is available \cite{kapoor2026guardrails}. An alternative is to keep the LLM assistance unrestricted and instead support users in deciding how much of the problem to solve themselves and how much to hand over to the assistant. However, design interventions that directly target these decisions are largely unexplored.

Here, we target two psychological mechanisms underlying users' decisions: (1) how they judge what their use of AI assistance means for their own learning and (2) how rewards shape the incentive to do the work themselves. The first mechanism is based on metacognition, which refers to how people assess and regulate their own thinking \cite{flavell1979metacognition}. Learners may delegate work without fully recognizing what this means for their own practice, especially when a fluent AI explanation creates a sense of understanding even though they have not applied the method themselves \cite{fan2025metacognitive}. The second mechanism is motivational incentives, as rewards can change the incentive to rely on external assistance rather than perform the work oneself. Prior research shows that making external assistance less rewarding or independent performance more rewarding can reduce offloading \cite{gilbert2020optimal, sachdeva2020reminders}. These two mechanisms later motivate our interventions based on metacognitive feedback that makes the implications of users' AI use visible, and an effort-based reward that encourages less extensive use of LLM assistance. We therefore ask the following research question:

\begin{quote}
\textbf{Research Question.} \emph{Can (a) metacognitive feedback or (b) an effort-based reward reduce answer offloading and improve subsequent unaided performance?}
\end{quote}

To answer this question, we design two interventions that target the offloading decision: (1) metacognitive feedback that makes the implications of the learner's LLM use visible, and (2) an effort-based reward that awards points for less extensive assistance. The metacognitive feedback is shown between learning items and describes how extensively a learner has requested assistance so far, what this implies for skill development, and prompts learners to reflect on their understanding in upcoming items. The reward changes the incentive for requesting assistance; i.e., a correct answer earns 10 points when the learner used no AI or only general guidance, 5 points when using an intermediate result, and 1 point when requesting the complete answer.

We test both interventions in a preregistered online experiment ($N = 704$) with a 2$\times$2 design and a no-AI control (Figure~\ref{fig:teaser}). Participants practiced fraction arithmetic with an LLM assistant. By default, the LLM assistant provided only guidance on how to approach the task and gave complete answers only on explicit request, thus leaving learners to decide how much of each problem to solve themselves or hand over to the assistant. The participants then completed a test without LLM assistance. We find that metacognitive feedback reduced answer offloading and improved test performance. We found no evidence that the reward affected either outcome. Our research makes the following contributions:

\begin{enumerate}
\item Empirically, we provide evidence that metacognitive feedback on a learner's AI requests reduces answer offloading and improves subsequent unaided performance. We found no evidence that a reward affected either outcome.
\item Theoretically, we extend the cognitive offloading paradigm to LLM-assisted learning by focusing on users' decisions about how much of a task to solve themselves and how much to hand over to the assistant.
\item Practically, we derive design recommendations for reducing the risk of deskilling without restricting LLM use. Importantly, our findings suggest that designers should make the implications of AI use visible, make complete-answer requests more deliberate, and support learning-oriented help-seeking.

\end{enumerate}

\section{Related Work}

\begin{table*}
  \caption{Research gaps across three main literature streams relevant to our work: AI-assisted learning, cognitive offloading, and metacognition.}
  \label{tab:streams}
{\footnotesize
\begin{tabularx}{\textwidth}{p{0.13\textwidth}p{0.53\textwidth}X}
\toprule
\textbf{Research stream} & \textbf{What prior work shows} & \textbf{Open question} \\
\midrule
\raggedright\textbf{AI assistance in learning} &
\textbf{AI assistance can support learning.} {AI} tutors can outperform in-class active learning \cite{kestin2025aitutor}. {AI}-generated hints can produce learning gains that do not differ from human-tutor hints \cite{pardos2024mathhints}.\par
\textbf{Unrestricted answer-giving harms unaided performance.} Unrestricted answer-giving can improve performance during practice while leading to weaker subsequent performance \cite{bastani2025guardrails, shen2026skillformation, liu2026persistence,rismanchian2026aleks}. Evidence further suggests that learning outcomes depend on how learners engage with the assistance \cite{lehmann2024classroom, liu2026persistence, gajos2022learning, shen2026skillformation}.\par
\textbf{Existing interventions mainly restrict the assistant.} Guardrails, Socratic tuning, and scaffolds change what the assistant will provide \cite{bastani2025guardrails, liu2024socraticlm, ma2025dbox, jurenka2024learnlm}. Such restriction can demotivate lower-proficiency learners or be circumvented \cite{myung2026scaffolding, kapoor2026guardrails}. &
Can interaction design support learners in deciding how much assistance to request without restricting the LLM assistance available to them? \\
\midrule
\raggedright\textbf{Cognitive offloading} &
\textbf{Offloading decisions depend on effort and incentives.} Using an external aid can reduce the cognitive effort required to complete a task \cite{risko2016offloading, gilbert2024valuebased}, and offloading behavior changes with the incentives attached to using that aid \cite{gilbert2020optimal}.\par
\textbf{Offloading is not harmful per se.} The consequences depend on what is handed over relative to the skill being developed \cite{risko2016offloading, kalyuga2025rethinking}. Offloading the cognitive work that a task is intended to build can undermine learning \cite{lodge2026offloading}.\par
\textbf{Offloading to LLM assistants is largely unexplored.} Existing work on offloading to AI is largely theoretical \cite{leondominguez2024risks, grinschgl2022distributed} or correlational \cite{gerlich2025aitools, lee2025criticalthinking}. Controlled studies show cognitive engagement can be preserved through the tailored design of human--AI interaction \cite{chen2025aidilemma, reicherts2025cognitivesupport, maier2026partnering}, but such scaffolds do not support users in regulating their own offloading decisions.
 &
Can changing the incentives around LLM assistance reduce answer offloading and improve unaided performance?\\
\midrule
\raggedright\textbf{Metacognition} &
\textbf{Assistance can distort judgments of learning.} Learners may perceive stronger understanding when support is available than their later independent performance warrants \cite{koriat2005illusions}. Students practicing with unrestricted LLM assistance did not recognize the subsequent performance decline they experienced \cite{bastani2025guardrails}.\par
\textbf{Learning with LLMs requires metacognitive regulation.} Users must plan their requests, evaluate responses, and adjust how they interact with the assistant \cite{tankelevitch2024metacognitive}, yet delegating cognitive work to an LLM assistant can reduce self-regulatory control \cite{fan2025metacognitive}.\par
\textbf{Metacognitive feedback shows promise for learning and AI use.} 
Metacognitive feedback can improve learning in computer-based
environments \cite{zheng2016srl, guo2022prompts}. In interactions with AI, existing interventions teach strategies for AI use \cite{garg2025prompttraining, valletorre2025chatbots, xiao2025pedagogicalprompting} or provide general guidance for reflecting during interaction  \cite{xu2025metacognitivesupport, singh2025metacogprompts}. To the best of our knowledge, no experiment has provided learners with feedback based on their own LLM use during learning.
 &
Can metacognitive feedback reduce answer offloading and improve unaided performance? \\
\bottomrule
\end{tabularx}
}
\end{table*}

\subsection{AI assistance in learning}

LLM assistants promise individualized support for learning at scale, for example by adapting explanations to the learner's needs, responding to solution attempts, and providing help when difficulties arise. A growing body of research demonstrates this potential. For example, AI tutors prompted to follow teaching best practices can outperform in-class active learning \cite{kestin2025aitutor}, AI-generated mathematics hints can produce learning gains comparable to human-tutor hints \cite{pardos2024mathhints}, and pedagogically designed assistants such as CodeAid already support entire programming courses \cite{kazemitabaar2024codeaid, liffiton2023codehelp}.  Beyond the assistant itself, {AI} can also personalize the learning path, for example by tracing a learner's evolving knowledge to recommend the next exercise~\cite{ozyurt2025exercise}.

However, there are concerns that LLM assistance during practice can also undermine skill acquisition \cite{kasneci2023chatgpt}. Bastani et al.~\cite{bastani2025guardrails} found that high-school students practicing mathematics with unrestricted LLM access performed better during practice but worse on a subsequent exam without LLM support than students in a no-AI control. Similar patterns have been observed, for example, in divergent and convergent creativity~\cite{kumar2025creativity}, coding~\cite{shen2026skillformation}, and reading comprehension and fraction problems~\cite{liu2026persistence}. Observational data from a large-scale learning platform points in a similar direction, namely that retention declined as students shifted work to generative AI~\cite{rismanchian2026aleks}.

Learning science provides one explanation for the above-mentioned pattern. Durable learning is known to depend on effortful processing, including  working through solution steps oneself (often referred to as ``desirable difficulties'') \cite{bjork2011difficulties}. When AI provides a worked solution, it removes effortful processing and leaves the learner with fewer opportunities to practice problem-solving independently~\cite{stadler2024cognitiveease}. The learner receives a worked solution without having constructed any part of it. Similar patterns were also observed in earlier tutoring systems; for example, learners who repeatedly requested hints until the answer was revealed tended to show lower learning gains~\cite{aleven2016help, baker2004gaming}. Together, these findings suggest that assistance can hinder learning when it replaces rather than supports the reasoning and problem solving the learner needs to practice.

Consequently, whether LLM assistance supports skill development may depend less on whether it is available than on how learners use it and how much of the problem solving they perform themselves. In line with this, the mere availability of LLM support shows no average effect, while solution-generating use is associated with less understanding and explanation-seeking use with more understanding \cite{lehmann2024classroom}. Similarly, interactions that prompt a learner's own thinking are known to preserve cognitive engagement better than those that deliver finished solutions \cite{xu2025productivereflective}, and explanations from an interactive LLM assistant can support learning more than having access to the answers alone \cite{kumar2025mathtutor}. AI-supported practice can also leave later performance without assistance intact \cite{bassner2025dissociation, kazemitabaar2023codegen}, whereas answer-giving AI can improve performance during practice while producing weaker subsequent learning \cite{bastani2025guardrails}. Consistent with this, a meta-analysis of coding studies finds that generative AI raises productivity during use but has no significant effect on learning, with in-session gains not transferring to unaided assessment~\cite{maier2026codingmeta}. This pattern echoes classical help-seeking research, which distinguishes instrumental requests for enough support to continue solving a problem oneself from executive requests that delegate the solution~\cite{nelsonlegall1981help}. As such, we expect that LLM assistance can have effects in both ways; i.e., either by scaffolding learners' practice or by replacing the cognitive work needed to develop the skill.


Existing interventions have mainly addressed these learning risks by changing what the assistant provides. For example, the use of guardrails in LLM assistants can restrict full answers and instead encourage the LLM to provide hints or guiding questions. Such guardrails were effective in a field experiment where they prevented the performance loss observed for students with unrestricted LLM access \cite{bastani2025guardrails}. Related designs steer assistants toward Socratic, pedagogical behavior or scaffold learners through problems step by step \cite{liu2024socraticlm, jurenka2024learnlm, ma2025dbox}. Similar approaches appear in commercial learning modes, including \emph{ChatGPT Study Mode}\footnote{\url{https://openai.com/index/chatgpt-study-mode/}}, \emph{Gemini Guided Learning}\footnote{\url{https://blog.google/products-and-platforms/products/education/guided-learning/}}, and \emph{Claude Learning Mode}\footnote{\url{https://www.anthropic.com/news/introducing-claude-for-education}}, which emphasize guiding questions and support for learners’ reasoning. However, restricting what the assistant can provide also has drawbacks. Withholding direct answers can demotivate lower-proficiency students and increase their reliance on the system \cite{myung2026scaffolding}, while users may also circumvent guardrails to access the complete answer \cite{kapoor2026guardrails}. However, these approaches all constrain the LLM assistant itself. Hence, whether careful interaction design can help learners decide how much help to request without restricting the LLM assistance available to them is unknown.

\subsection{Cognitive offloading}

Cognitive offloading refers to reducing cognitive effort by shifting a part of a task onto an external aid, for example by writing something down or setting a reminder instead of keeping an intention in mind \cite{risko2016offloading}. A specific example is the decision of how much help to request from an LLM assistant, which is the focus of our paper. HCI has long examined how cognitive work is distributed across people and external artifacts \cite{hollan2000distributed}; however, LLMs change the scope of what can be offloaded. In particular, a learner can ask for a hint and continue working through the task, or request the complete solution and hand much of the cognitive work to the assistant. The amount of help a learner requests can therefore be understood as an offloading decision.

Offloading is not harmful per se. It can free cognitive capacity by shifting work that is not central to the current goal onto an external aid and can even improve memory for new materials \cite{risko2016offloading, storm2015saving}. Whether the reduced cognitive effort is helpful or harmful for skill development therefore depends on what is offloaded by the learner \cite{kalyuga2025rethinking}. A recent synthesis in education makes this distinction explicit by separating beneficial offloading from offloading cognitive work  that a task is intended to build \cite{lodge2026offloading}. For example, asking how to approach a problem leaves the actual practice intact, whereas requesting the complete solution removes much of the practice itself; at the extreme, users may even adopt AI output without engaging with the underlying task \cite{shaw2026surrender}. We refer to this negative form as \emph{answer offloading}.

Whether people offload depends on the perceived benefit relative to the effort of using an external aid \cite{gilbert2024valuebased}, as well as on their confidence in their own ability \cite{hu2019metamemory}. For learners, this can favor offloading: when the immediate goal is to complete the task, delegating the problem-solving saves effort, while doing the work oneself offers little immediate reward. At the same time, learners may not fully recognize what they give up by doing so, especially because fluent assistance can inflate their perceived learning success even when they have done less of the cognitive work themselves \cite{koriat2005illusions}, while reliance on the assistant can reduce the self-regulation involved in learning \cite{fan2025metacognitive} (see the discussion on metacognition in the next section). Together, these mechanisms point to two ways of reducing detrimental offloading: (1) changing the incentives for doing the work oneself and (2) helping learners recognize the potentially negative implications of AI use.

In human--AI interaction, \emph{reliance} describes the extent to which users adopt AI output in their own judgment \cite{schemmer2023appropriate, raees2026reliancereview}. Much of this literature asks how users should rely on AI when its advice may be wrong. In learning, however, even correct assistance can create a different challenge: users may rely on the assistant in ways that reduce their own cognitive engagement. The HCI community has begun to frame generative AI through the lens of cognitive offloading \cite{tankelevitch2025toolsforthought}, even though work on offloading to AI assistants is largely theoretical \cite{leondominguez2024risks, grinschgl2022distributed} or correlational \cite{gerlich2025aitools, lee2025criticalthinking}. Controlled studies are largely the exception. For example, grading the extent of AI assistance can preserve more cognitive engagement than full automation \cite{chen2025aidilemma}, while, in a creativity task, an LLM that asks questions to elicit a user's own ideation process rather than automating the idea generation can preserve idea quality and perceived ownership at the price of higher perceived effort \cite{maier2026partnering}. However, both exceptions redesign how the assistant responds; in contrast, whether users can instead be supported in regulating how much cognitive work they hand over remains unknown.


\subsection{Metacognition}
\label{sec:metacognition}

Metacognition refers to thinking about one's own thinking, which is classically divided into monitoring one's cognitive states and controlling one's cognitive activities \cite{flavell1979metacognition, nelson1990metamemory}. Deciding how much work to hand over to an external aid is one such control decision, and it depends on how learners assess their own understanding and need for assistance.

Using LLMs already requires metacognitive monitoring and control. Because users largely steer the interaction themselves, they must decide what to ask, monitor whether the responses serve their goals, and adjust their interaction accordingly \cite{tankelevitch2024metacognitive}. These strategies can thus influence performance with AI. In a field experiment with consultants, LLM access improved creative performance particularly for consultants who used such metacognitive strategies \cite{sun2025creativity}. Learning adds another challenge for metacognitive demands, as learners must judge not only whether the assistance helps them complete a task, but also whether they learn to perform the task independently as a result. Importantly, learners tend to overestimate their understanding when they practice a task with support but where such support is unavailable during the test \cite{koriat2005illusions}. One reason is that such LLM support may make learners feel more fluent than they actually are. For instance, students who practiced with unrestricted LLM access did not recognize the later performance decline during an unaided exam \cite{bastani2025guardrails}; this finding is consistent with broader concerns that AI assistance may reduce self-regulation during learning \cite{fan2025metacognitive}. A learner working with an LLM assistant needs to monitor their own understanding and regulate how they use the assistant in response. 

Metacognitive feedback offers a principled way to help users in both monitoring their understanding and regulating their use of AI assistance. In AI-assisted decision making, tailored interventions (e.g., cognitive forcing functions, prompts to encourage a user's own reasoning, and explanations that facilitate verification) can help users evaluate AI advice more critically \cite{bucinca2021trust, danry2023askme, vasconcelos2023explanations}. In computer-based learning environments, metacognitive prompts help learners monitor their understanding \cite{bannert2015prompts}, while metacognitive feedback can improve learning outcomes, with meta-analyses reporting moderate effects \cite{zheng2016srl, guo2022prompts}. More recently, work has extended metacognitive feedback to generative AI interactions. For example, support agents can encourage users to reflect on their goals and reasoning while working with generative AI \cite{gmeiner2025metacog}. In educational contexts, interventions teach prompting and help-seeking strategies through separate training activities \cite{xiao2025pedagogicalprompting, valletorre2025chatbots}, or provide general metacognitive feedback while students work with the tool \cite{xu2025metacognitivesupport, singh2025metacogprompts}. To the best of our knowledge, no prior work has provided learners with feedback based on their own AI use during learning and whether such feedback can improve self-regulation is thus unclear.

\subsection{Research gap}
In summary, prior research (Table~\ref{tab:streams}) shows that LLM assistance can reduce cognitive engagement and thus hinder skill development. However, no study has proposed and evaluated interaction design interventions that support users in regulating their own AI use to improve subsequent unaided performance. Hence, reducing answer offloading through interaction design remains an open challenge.

\section{Hypotheses}

We develop and experimentally test different interventions to reduce answer offloading and improve skill acquisition. We derive three preregistered hypotheses with two behavioral outcomes, namely, answer offloading during learning and unaided performance during subsequent tests (Section~\ref{sec:measures}). The hypotheses and analysis plan were preregistered on AsPredicted (\url{https://aspredicted.org/gx2fa8.pdf}).

\subsection{\factorAssistant Access to AI}

AI support can oftentimes improve immediate task performance \cite[e.g.,][]{gajos2022learning}, but it may also reduce critical thinking and the processing depth that people invest in solving tasks \cite{stadler2024cognitiveease, lee2025criticalthinking, chen2025aidilemma}. Given that learning requires such effortful processing \cite{bjork2011difficulties, soderstrom2015learning}, better performance during practice with LLM assistance does not necessarily translate into actual skill acquisition. Consistent with this, studies across various domains have found lower unaided test performance after practice with unrestricted LLM assistance compared to an unassisted control \cite{bastani2025guardrails, liu2026persistence, shen2026skillformation}. We thus expect:

\begin{description}
  \item[H1:] \emph{Participants with LLM assistance during learning perform worse on the subsequent unaided test than participants who learn without LLM assistance.}
\end{description}

\subsection{\factorFeedback Metacognitive feedback}

Offloading decisions depend in part on how learners assess their own understanding and need for assistance. LLM support can make this assessment more difficult, as fluent answers may increase perceived understanding while reducing self-regulatory engagement \cite{koriat2005illusions, fan2025metacognitive}. 
Interventions based on metacognition have been found to improve learning behavior and student performance \cite{zheng2016srl, guo2022prompts}. Recent work further suggests a similar benefit for self-regulated learning with AI \cite{xu2025metacognitivesupport, lee2025metacogcalibration}. For example, in the context of intention offloading, metacognitive feedback (e.g., via strategic advice \cite{gilbert2020optimal} and brief training with feedback \cite{ngai2026metacognitive}) can help people regulate when to rely on external aids. Transferring these results to our context of LLM-assisted learning, we expect:

\begin{description}
  \item[H2a:] \emph{Participants who receive metacognitive feedback are less likely to offload the answer on a learning item than participants without metacognitive feedback.}
  \item[H2b:] \emph{Participants who receive metacognitive feedback perform better on the subsequent unaided performance test than participants without metacognitive feedback.}
\end{description}

\subsection{\factorReward Effort-based reward}

Cognitive effort depends partly on the incentives attached to doing the work oneself versus relying on an external aid \cite{gilbert2024valuebased, kool2018mentallabour}. Hence, reducing the incentives for cognitive offloading should shift learners toward solving more of the task themselves. For example, in reminder-setting paradigms, people offload less when they earn fewer points for using an external reminder than recalling the item unaided \cite{gilbert2020optimal}. Extending this idea to LLM assistance, a reward that awards fewer points after more extensive assistance should make doing more of the work oneself more attractive and should reduce answer offloading. We therefore expect rewards to reduce answer offloading and thus improve learning. 
Consequently, we hypothesize:

\begin{description}
  \item[H3a:] \emph{Participants who receive an effort-based reward are less likely to offload the answer on a learning item than participants without an effort-based reward.}
  \item[H3b:] \emph{Participants who receive an effort-based reward perform better on the subsequent unaided performance test than participants without an effort-based reward.}
\end{description}

\section{Method}
We conducted a preregistered online experiment in August 2026 in which participants practiced fraction arithmetic with an LLM assistant and afterwards completed a test without LLM assistance (which we refer to as ``unaided performance''). We preregistered the hypotheses, design, sample size, exclusion rules, and analysis plan on AsPredicted before data collection. Analysis code, anonymized data, and LLM reporting following best-practice guidelines \cite{feuerriegel2026checklist} are available in our anonymized repository.\footnote{\url{https://anonymous.4open.science/r/cognitive-offloading-CHI2027}}

\subsection{Study procedure}
\label{sec:procedure}

Our experiment comprised six steps. (1) Participants were redirected from Prolific to our custom study platform and were randomly assigned to one of the five conditions. (2) A baseline block collected demographics and other controls such as AI use frequency (Section~\ref{sec:measures}). (3) Participants received a brief review of the basic rules needed to solve fraction arithmetic problems, and participants in conditions with assistant access were additionally told that the LLM assistant would be available during the learning phase, but that the subsequent test measures unaided performance. (4) Participants then completed 10 learning items, with the correct solution shown after each item. The first exposure to the intervention is after the practice item; accordingly, we measure answer offloading not on the first item but only on the subsequent nine items. Depending on condition, participants completed the learning phase without LLM assistance (no-AI control), with LLM assistance but no intervention (AI-only), with metacognitive feedback, with reward, or with both interventions (Section~\ref{sec:intervention-groups}). (5) After the learning block, participants rated their perceived mental effort and predicted how many of the six test items they would solve correctly. (6) Participants completed the six test items without LLM assistance (i.e., unaided performance) and rated their confidence after each one as well as their perceived mental effort during the test. Finally, participants completed an honesty check and were briefly debriefed about the study background.

\subsection{Learning platform}
\label{sec:technical-design}

\subsubsection{System architecture}
\begin{figure*}
  \centering
  \includegraphics[width=\textwidth]{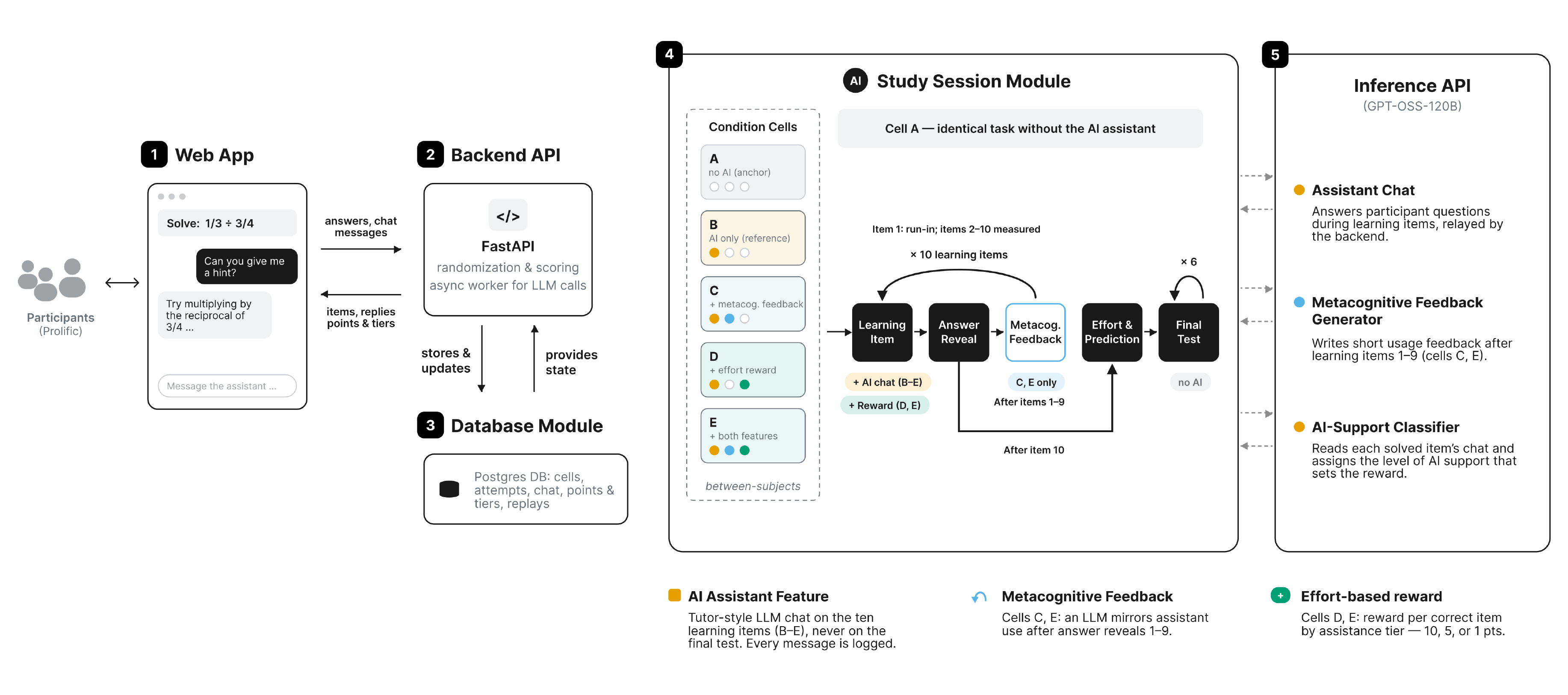}
  \caption{Study system architecture. Participants interact with the web app~(1), which communicates with the backend API~(2) and database~(3). The study session module~(4) manages the item sequence for the corresponding intervention. The LLM assistant (together with the intervention-specific extensions) accesses a state-of-the-art open-weight LLM~(5).}
  \Description{System architecture arranged left to right in five blocks. At the far left, participants recruited on Prolific interact with the web app (1), a phone-style mockup showing the item ``Solve: 1/3 divided by 3/4'', a learner message ``Can you give me a hint?'', and the assistant's reply ``Try multiplying by the reciprocal of 3/4''. Arrows connect the web app to the backend API (2): answers and chat messages flow to the backend, and items, replies, points, and tiers flow back. The backend runs FastAPI and handles randomization, scoring, and an asynchronous worker for LLM calls. It stores and updates records in, and reads study state from, the database module (3), a Postgres database holding cells, attempts, chat, points and tiers, and replays. The study session module (4) contains a between-subjects panel of five condition cells: A, no AI (anchor); B, AI only (reference); C, plus metacognitive feedback; D, plus the effort-based reward; and E, plus both features. A banner notes that cell A does the identical task without the AI assistant. The session flow runs Learning Item, Answer Reveal, Metacognitive Feedback, Effort and Prediction, and Final Test, with a loop of ten learning items (item 1 practice, items 2 to 10 measured) and a final-test loop of six items; tags mark AI chat in cells B to E and reward in cells D and E, metacognitive feedback after items 1 to 9 in cells C and E only, and no AI on the final test. Dashed arrows link the session module to the inference API (5), serving GPT-OSS-120B, which provides three services: the assistant chat that answers questions during learning items, the metacognitive feedback generator that writes short usage feedback after learning items 1 to 9 in cells C and E, and the AI-support classifier that reads each solved item's chat and assigns the assistance level that sets the reward. A bottom legend defines the AI assistant feature (tutor-style LLM chat on the ten learning items in cells B to E, never on the final test, with every message logged), the metacognitive feedback (cells C and E, an LLM mirrors assistant use after answer reveals 1 to 9), and the effort-based reward (cells D and E, points per correct item by assistance tier: 10, 5, or 1).}
  \label{fig:architecture}
\end{figure*}

We built a custom web platform so that every part of the participant experience (item presentation, chat, scoring, intervention delivery) was under experimental control and fully logged. For this we used a React-based frontend, which communicates with a Python backend that manages participants' sessions, with data stored in a PostgreSQL database. We built the LLM assistant around GPT-OSS-120B \citep{openai2025gptoss}, which is an open-weight, frontier LLM and which we accessed through Amazon Bedrock (\texttt{openai.gpt-oss-120b-1:0}). We set the temperature to 0 for more reproducible behavior. During the learning phase, the assistant appeared in a chat panel shown next to the current item. Each item has its own conversation. Finally, all chat transcripts, answers, timestamps, and intervention exposures are stored server-side. Figure~\ref{fig:architecture} shows the architecture.

\subsubsection{LLM assistant}
Following best practices for validating and reporting LLM-based systems in research \cite{felixnavarro2026reporting, lin2026validity}, we designed and validated the system prompt of the LLM assistant before study deployment. The system prompt holds the fixed instructions that govern the assistant's behavior. We wrote these instructions so that offloading would reflect an active, deliberate choice, since, by default, an LLM supplies the complete answer even when it was not requested. We therefore developed the prompt iteratively against a fixed evaluation set of 55 synthetic participant messages that spanned the request types we anticipated. For each revision, we generated replies through the same chat pipeline that the study later used, and the study's classifier labeled the level of assistance each reply supplied. In addition, we checked every number in a reply against the item's verified solution. We revised the prompt and the generation settings (i.e., the temperature and the maximum reply length) until three complete passes over all 55 messages showed the intended behavior. In these passes, no reply stated a solution specific to the current item that the participant had not requested, the assistant never withheld an explicitly requested answer, and every stated number was correct. We froze this configuration for the study. The full system prompt is provided in Appendix~\ref{app:system-prompt}. Details about the above evaluation are available in our repository. We provide representative participant--assistant dialogues in Appendix~\ref{app:transcripts}.

\subsubsection{Classifying extent of LLM support for reward computation}
\label{sec:assistance-classification}

To measure the extent of answer offloading and subsequently compute corresponding rewards, we classified how much LLM support participants requested on each learning item. For this, we built a custom LLM-based classifier (\emph{offloading classifier}). We distinguished four levels: (i)~no AI use, (ii)~guided help, (iii)~an intermediate result, and (iv)~the complete answer. If participants interacted with the LLM multiple times on the same learning item, the item was assigned the highest level of support requested across all messages.

In our implementation, we combined (a)~an LLM-based classifier with (b)~a rule-based check of numerical values in the conversation. These are as follows: The numerical check parsed every number in the conversation, distinguishing a new item-specific value from confirmation of one the participant had already produced. The LLM-based classifier (GPT-OSS-120B, temperature~0) then classified each request using a fixed rubric; for example, any request for the value of the full expression was classified as a complete answer, regardless of how it was phrased. We developed the rubric using 198 synthetic tutoring dialogues covering clear, boundary, and adversarial cases. On a separate set of 198 dialogues held out from this development and never used for tuning, the classifier agreed with the intended labels in every case, yielding no confirmed classification error. We froze the classifier configuration before data collection. The classifier prompt and evaluation details are provided in Appendix~\ref{app:classifier-prompt} and our repository.

\subsubsection{Task materials}

Following \citet{liu2026persistence}, we use fraction arithmetic as the task domain. Fractions suit our purpose for three reasons: (1) answers are unambiguously right or wrong, (2) the difficulty can be controlled by construction rather than estimated from pilot data, and (3) competence is learnable, so offloading effects can be observed within a single session. 

The items for learning and the subsequent performance were designed to assess the same underlying fraction arithmetic while avoiding simple recall of earlier problems. We focus on different operations, including addition, subtraction, multiplication, and division. The items from the unaided test used new operands and expressions but required the same operations practiced during learning. In this way, the subsequent test assessed transfer of the practiced procedures rather than memory for earlier items or answers. All items were generated from predefined expression templates spanning one to three arithmetic steps. Details about the different fraction problems are provided in Appendix~\ref{app:item-bank}.

\begin{figure*}
  \centering
  \includegraphics[width=\textwidth]{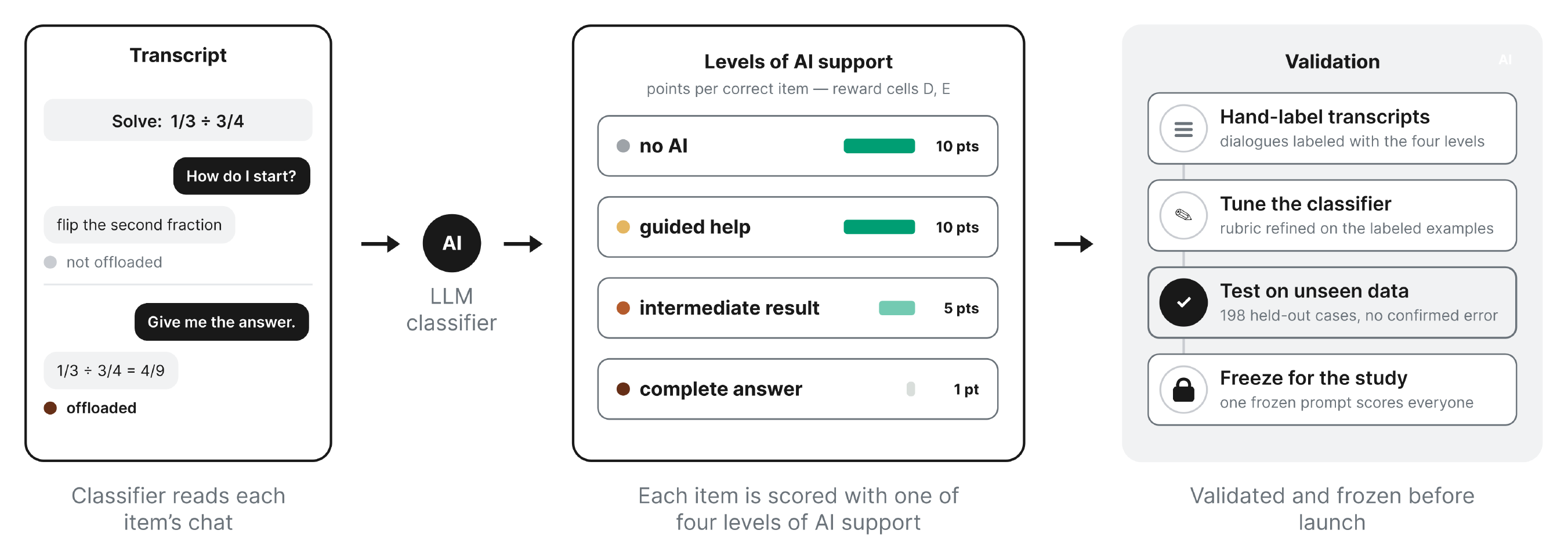}
  \caption{\factorReward Classifying extent of LLM support for reward computation. An LLM classifier uses the chat from each learning item (left) to label the highest level of LLM support requested. This is used in the reward conditions to determine how many points a correct answer earns. We distinguished four levels: (i)~no AI use, (ii)~guided help, (iii)~an intermediate result, and (iv)~the complete answer. The classifier was developed on 198 synthetic dialogues, evaluated on a separate set of 198 dialogues held out from development, and frozen for the study before launch (right).}
  \Description{Three-panel diagram. Left panel: an example chat transcript for the item 1/3 divided by 3/4, in which a first request (``How do I start?'') receives guidance and is marked as not offloaded, while a second request (``Give me the answer.'') receives the full result and is marked as offloaded. Center panel: the four assistance tiers (no AI use, guided help, intermediate result, complete answer) with the points a correct item earns in the reward conditions (10, 10, 5, and 1). Right panel: the validation pipeline in four steps: labeling synthetic dialogues with the four assistance levels, tuning the classifier rubric, testing on a separate set of 198 dialogues held out from development with no confirmed classification error, and freezing the prompt for the study.}
  \label{fig:classifier}
\end{figure*}

\subsection{Interventions}
\label{sec:intervention-groups}

Participants were randomly assigned to one of five between-subjects conditions in a 2$\times$2-factorial-plus-control design: (a)~a no-AI control that includes the same items as in the other conditions but without access to an LLM assistant; (b)~an AI-only condition, in which the LLM assistant was available without either intervention, thus representing unrestricted use without a tailored interaction design; (c)~a metacognitive feedback condition, in which participants received feedback on how they had used the LLM assistant and what this meant for their own skill development; (d)~an effort-based reward condition, in which correct answers earned fewer points when participants requested more extensive LLM assistance; and (e)~a combined condition, in which participants received both the metacognitive feedback and the reward. We provide details on how we operationalize the two main interventions below. The instructions and feedback texts shown to participants are reported in Appendix~\ref{app:intervention-texts}.

\subsubsection{\factorFeedback Metacognitive feedback}

The metacognitive feedback targets how learners assess their own use of LLM assistance when deciding how much help to request. This is motivated by the fact that learners oftentimes fail to recognize how much cognitive work they hand over to the assistant or what this means for their own practice (Section~\ref{sec:metacognition}), so the feedback makes one's own LLM use visible before making the next offloading decision.

\begin{figure*}
  \centering
  \includegraphics[width=\textwidth]{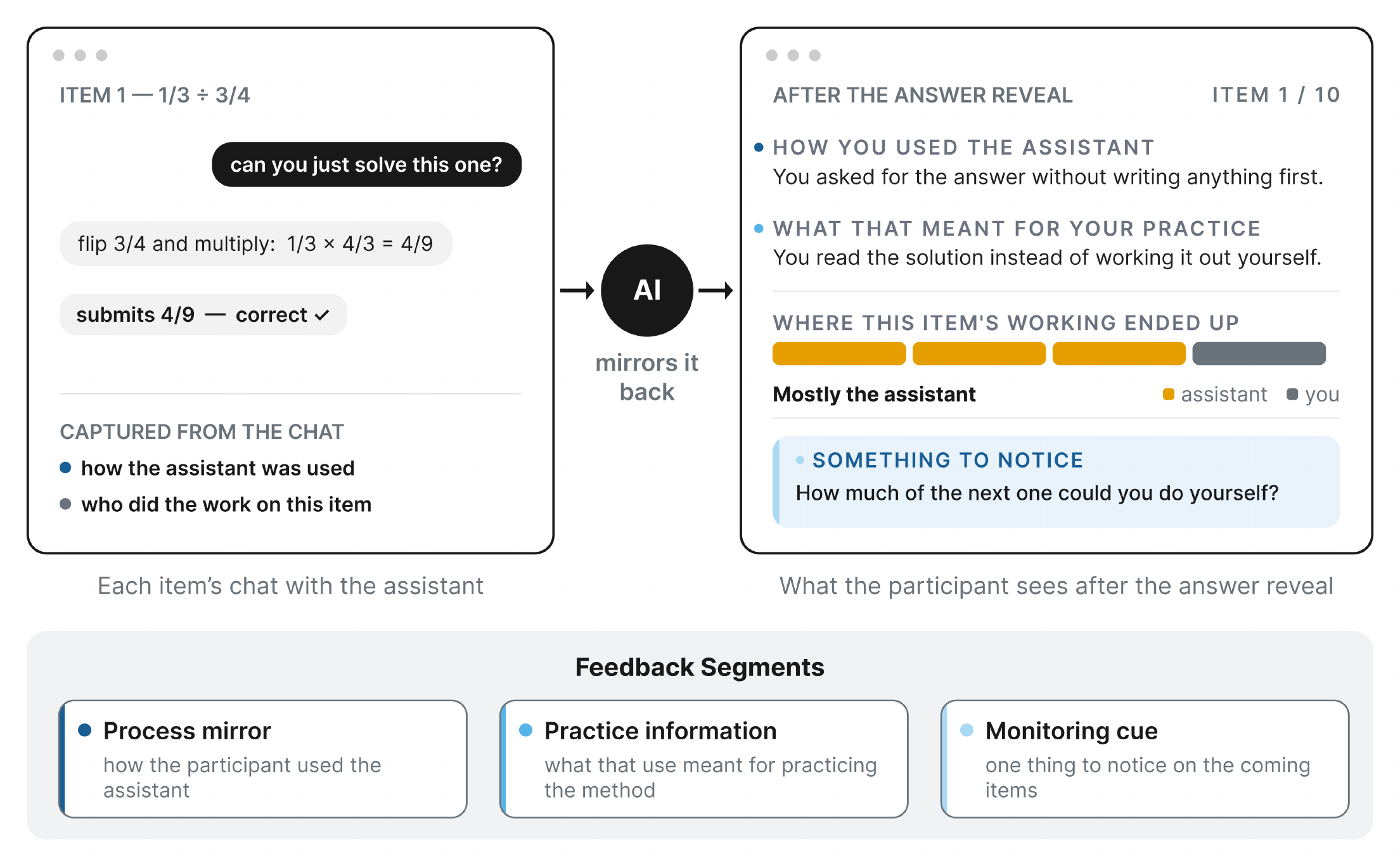}
  \caption{\factorFeedback Metacognitive feedback. To provide metacognitive feedback, an LLM uses the participant's chat history  up to the current item (left, showing the current item) to generate brief feedback, which is shown after revealing the correct answer (center). The metacognitive feedback contains three parts (right): a process mirror (``How you used the assistant''), practice information (``What that meant for your practice''), and a monitoring cue (``Something to notice''). We further show a bar chart (``Where this item's working ended up'') that places the item on a four-step axis from ``All with you'' to ``Mostly the assistant''.}
  \Description{Two large panels connected by an AI symbol and arrows, with three feedback-segment cards below. The left panel shows item 1, 1/3 divided by 3/4. The participant asks ``can you just solve this one?'', the assistant replies ``flip 3/4 and multiply: 1/3 times 4/3 = 4/9'', and the participant submits 4/9, marked correct. Below the chat, two bullets identify what is captured: how the assistant was used and who did the work on this item. The right panel shows the feedback screen after the answer reveal, labeled item 1 of 10. Under ``How you used the assistant'', it states ``You asked for the answer without writing anything first.'' Under ``What that meant for your practice'', it states ``You read the solution instead of working it out yourself.'' Under ``Where this item's working ended up'', a four-segment bar has three gold segments for the assistant and one gray segment for the participant, summarized as ``Mostly the assistant''. A blue box headed ``Something to notice'' reads ``How much of the next one could you do yourself?'' The bottom cards define the process mirror as how the participant used the assistant, practice information as what that use meant for practicing the method, and the monitoring cue as one thing to notice on the coming items.}
  \label{fig:feedback-generator}
\end{figure*}

The metacognitive feedback appeared after each learning item, between revealing the answer and the next item (see Figure~\ref{fig:feedback-generator}). For each item, an LLM (GPT-OSS-120B, temperature~0.7) generated the feedback based on the participant's stored learning-phase record (see Appendix~\ref{app:feedback-prompt}). Each feedback consists of three brief parts: a \emph{process mirror} reflecting how the participant used the assistant, \emph{practice information} explaining what this meant for their own practice, and a \emph{monitoring cue} directing attention to their understanding on upcoming items.

The design of the three parts follows theory on self-regulated learning, in which learners first monitor how they work and then adjust their behavior in response \cite{nelson1990metamemory, butler1995feedback, kluger1996feedback, hattie2007feedback}. The practice information builds on the finding that reading a solution can feel like understanding it while providing less opportunity to perform the cognitive work required for learning \cite{koriat2005illusions, soderstrom2015learning}. The monitoring cue draws on prior work showing that metacognitive prompts can direct learners' attention to their own understanding during learning \cite{bannert2015prompts}. Importantly, the feedback was informational (and not instructional; i.e., without telling participants how much assistance to request in the next item), so any behavior change reflected the participant's own regulation.

\subsubsection{\factorReward Effort-based reward}

The reward used non-monetary points, which are displayed after every item together with a running total. In the reward condition, the points for a correct learning item depended on the highest level of LLM support requested on that item (see Figure~\ref{fig:classifier}): 10 points with no LLM use or only requests for guidance, 5 points after requesting an intermediate result, and 1 point after requesting the complete answer. Thus, any request for the complete answer reduced the reward for that item to 1 point, whereas any number of guidance requests preserved the full 10 points.

Outside the reward conditions, participants still received points as an incentive for correct performance: every correct answer earned 10 points and every incorrect answer 0 points. However, these points were independent of how participants used the LLM assistant. Only in the reward and combined conditions did the number of points awarded for a correct answer vary with the level of LLM support requested.

\subsection{Measures} 
\label{sec:measures}


$\bullet$\,Our primary outcomes are: \textbf{Answer offloading} (H2a, H3a) is the proportion of the nine learning items on which the participant requested the complete answer from the assistant. Answer offloading is derived from the requested level of LLM support (see Section~\ref{sec:assistance-classification}). As preregistered, the practice item is excluded from the answer offloading outcome because the feedback and reward manipulations first take effect for the second item. \textbf{Unaided performance} (H1, H2b, H3b) is the proportion of the six final-test items answered correctly. 

$\bullet$\,We measured the following exploratory outcomes. First, we assessed two aspects of metacognitive judgment. \emph{Metacognitive accuracy} compares the actual minus predicted test score, with positive values indicating underestimation. \emph{Metacognitive sensitivity} captures how well a participant's confidence distinguishes between their correct and incorrect answers. Following \citet{fernandes2026metacognition}, we calculate this at the participant level using the area under the receiver operating characteristic curve (AUC), relating confidence ratings to item correctness. Higher values indicate that the participant tended to give higher confidence ratings to answers that were correct than to answers that were incorrect. An AUC of $0.5$ indicates chance-level discrimination, whereas an AUC of $1$ indicates perfect discrimination between correct and incorrect responses.

We further measured: \textbf{trust in AI}, adapted from the six-item propensity-to-trust scale of \citet{merritt2013} (``My tendency to trust AI is high.''; Cronbach's $\alpha = 0.94$); \textbf{need for cognition} with the six-item short form of the Need for Cognition Scale \citep{coelho2020} (``I would prefer complex to simple problems.''; $\alpha = 0.90$); \textbf{AI use frequency} with a single item (``How often do you use AI assistants (like ChatGPT) for work or study tasks?''); and \textbf{perceived confidence} with a single item adapted from \citet{lee2025criticalthinking} (``How confident are you in your ability to do this task without GenAI?''). \textbf{Perceived mental effort} was measured with the single-item 9-point scale of \citet{paas1992} (``How much mental effort did those problems require?''), once after the learning block and once after the final test. Appendix~\ref{app:overview-measures} shows all constructs with their sources, item counts, and response formats.

\subsection{Participants}
\label{sec:sample}
Before preregistration, we determined the required sample size based on  a one-sided two-sample $t$-test comparing the AI-only and control conditions, with $\alpha = .05$, $1 - \beta = 0.80$, and a smallest effect size of interest of $d = 0.30$. This yielded a minimum required sample of 695 participants across the five conditions. 

Participants were eligible if they had registered on Prolific before August 2026, resided in the UK, were at least 18 years old, reported English as their first language, and had completed at least 50 previous submissions with an approval rate of at least 99\%. Participation was restricted to desktop devices. To account for attrition, we recruited 720 participants, of whom five did not complete the study. We excluded 11 due to the honesty check. 
Thus, the final sample comprised $N = 704$ participants, aged 18 to 80 years ($M = 39.1$, $SD = 12.8$). Of these, 375 (53.3\%) identified as male, 326 (46.3\%) as female, and 3 (0.4\%) as another gender.

\subsection{Ethics}
\label{sec:ethics}
Our study received IRB approval from the ethics committee at LMU Munich School of Management (ETH-SOM-089; approved July 31, 2026). All participants provided their informed consent and could withdraw at any time. Participants were paid independent of their performance, meeting Prolific's fair-pay policy. Data were collected and stored in accordance with local data privacy laws.

\subsection{Statistical analysis}
\label{sec:statistical-analysis}
Following our preregistered analysis plan, all confirmatory tests were one-sided at $\alpha= .05$. Exploratory analyses used two-sided tests. Each estimate is accompanied by a two-sided 95\% confidence interval (CI). Effect sizes are expressed as odds ratios. We used item-level mixed-effects logistic regression models for all primary confirmatory analyses, fitted with the \texttt{lme4} package \cite{bates2015lme4}. This approach estimates the binary outcome and keeps each trial as its own observation while accounting for the fact that trials from the same participant are correlated. As a robustness check, we also report a participant-level $t$-test and ANOVAs in Appendix~\ref{app:robustness} (Table~\ref{tab:hypotheses-robustness}), which led to the same conclusions. All analyses were conducted using \emph{R} 4.6.0. 

\section{Results}

\begin{table*}[t]
  \caption{Descriptive statistics by condition for the primary outcomes, AI use, learning-phase performance, and perceived mental effort.}
  \label{tab:descriptives}
  \Description{Table reporting, for each of the five conditions, the number of participants; the means and standard deviations of complete-answer offloading, unaided final-test performance, and learning performance; the number and percentage of participants who used the assistant at least once and who requested a complete answer at least once; and the means and standard deviations of self-reported mental effort during the learning phase and during the final test. The answer-offloading and unaided-performance columns are shaded.}
  \vspace{-0.3cm}
  \small
  \setlength{\tabcolsep}{4pt}
  \setlength{\aboverulesep}{0pt}
  \setlength{\belowrulesep}{0pt}
  \renewcommand{\arraystretch}{1.2}
  \begin{tabular}{l c YY YY cc c c cc cc}
    \toprule
    & & \multicolumn{2}{Y}{Answer}
      & \multicolumn{2}{Y}{Unaided}
      & \multicolumn{2}{c}{Performance}
      & \multicolumn{1}{c}{Used AI}
      & \multicolumn{1}{c}{Offloaded}
      & \multicolumn{2}{c}{Mental effort}
      & \multicolumn{2}{c}{Mental effort} \\
    & & \multicolumn{2}{Y}{offloading}
      & \multicolumn{2}{Y}{performance}
      & \multicolumn{2}{c}{(learning phase)}
      & \multicolumn{1}{c}{$\geq1$ item}
      & \multicolumn{1}{c}{$\geq1$ item}
      & \multicolumn{2}{c}{(learning phase)}
      & \multicolumn{2}{c}{(final test)} \\
    \cmidrule(lr){3-4} \cmidrule(lr){5-6} \cmidrule(lr){7-8}
    \cmidrule(lr){9-9} \cmidrule(lr){10-10}
    \cmidrule(lr){11-12} \cmidrule(lr){13-14}
    Condition & \textit{N}
    & \textit{M} & \textit{SD} & \textit{M} & \textit{SD} & \textit{M} & \textit{SD}
    & \textit{n} (\%) & \textit{n} (\%)
    & \textit{M} & \textit{SD} & \textit{M} & \textit{SD} \\
    \midrule
    \condDots{0}{0}{0}No-AI control & 146 & --- \textsuperscript{a} & --- & 0.65 & 0.32 & 0.68 & 0.28 & --- \textsuperscript{a} & --- \textsuperscript{a} & 7.63 & 1.47 & 7.57 & 1.60 \\
    \condDots{1}{0}{0}AI-only       & 137 & 0.18 & 0.30 & 0.62 & 0.34 & 0.77 & 0.24 & 94 (68.6\%) & 54 (39.4\%) & 7.55 & 1.47 & 7.77 & 1.45 \\
    \condDots{1}{1}{0}M. feedback   & 144 & 0.13 & 0.22 & 0.67 & 0.30 & 0.79 & 0.20 & 96 (66.7\%) & 55 (38.2\%) & 7.24 & 1.71 & 7.52 & 1.57 \\
    \condDots{1}{0}{1}Effort reward & 138 & 0.15 & 0.24 & 0.58 & 0.32 & 0.74 & 0.25 & 84 (60.9\%) & 57 (41.3\%) & 7.84 & 1.43 & 8.09 & 1.23 \\
    \condDots{1}{1}{1}Combined      & 139 & 0.10 & 0.18 & 0.64 & 0.34 & 0.78 & 0.23 & 73 (52.5\%) & 46 (33.1\%) & 7.51 & 1.66 & 7.70 & 1.49 \\
    \bottomrule
    \multicolumn{14}{@{}p{\dimexpr\textwidth-2\tabcolsep\relax}@{}}{%
      \footnotesize
      Answer offloading and learning performance are proportions of the nine measured learning items; unaided performance is the proportion of the six final-test items. The two AI-use columns give the number and percentage of participants who used the AI assistant at least once and who offloaded at least one answer. Mental effort is a single-item self-report on a nine-point scale, collected after the learning phase and after the final test.
      \textsuperscript{a}Not applicable in the control, because no LLM assistant was available.} \\
  \end{tabular}
\end{table*}

\subsection{Main results}

\begin{table*}[t]
  \caption{Confirmatory hypothesis tests using item-level binomial mixed-effects logistic regressions. Odds ratios below 1 indicate lower odds of the outcome. Arrows give the direction of each preregistered prediction; $p$-values are one-sided in that direction and confidence intervals are two-sided. Observations are item-level responses, reported alongside the
    number of participants.
    $^{*}p<.05$, $^{**}p<.01$, $^{***}p<.001$.}
  \label{tab:hypotheses}
  \Description{Table reporting, for each of the five preregistered hypotheses, the
    outcome tested, the predicted direction, the odds ratio with its 95 percent
    confidence interval, the one-sided p-value, the number of observations and
    participants, and whether the hypothesis was supported. Rows are grouped by
    experimental factor: AI access, metacognitive feedback, and effort-based reward.}
  \vspace{-0.3cm}
  \begin{tabular}{@{}llcccc@{}}
    \toprule
    Hyp. & Outcome & OR [95\% CI] & $p$ & Obs. (\#Participants) & Supported? \\
    \midrule
    \multicolumn{6}{l}{\factorAssistant\textit{AI access vs.\ no-AI control}}\\
    H1  & Unaided performance ($\downarrow$)  & $0.80\;[0.42,\,1.51]$ & $.244$    & $1{,}698\;(283)$ & \textcolor{BrickRed}{\ding{55}} \\
    \midrule
    \multicolumn{6}{l}{\factorFeedback\textit{Metacognitive feedback}}\\
    H2a & Answer offloading ($\downarrow$) & $0.47\;[0.22,\,1.00]$ & $.026^{*}$ & $5{,}022\;(558)$ & \textcolor{ForestGreen}{\ding{51}} \\
    H2b & Unaided performance ($\uparrow$) & $1.51\;[0.98,\,2.33]$ & $.030^{*}$ & $3{,}348\;(558)$ & \textcolor{ForestGreen}{\ding{51}} \\
    \midrule
    \multicolumn{6}{l}{\factorReward\textit{Effort-based reward}}\\
    H3a & Answer offloading ($\downarrow$) & $0.66\;[0.31,\,1.40]$ & $.139$    & $5{,}022\;(558)$ & \textcolor{BrickRed}{\ding{55}} \\
    H3b & Unaided performance ($\uparrow$) & $0.76\;[0.50,\,1.18]$ & $.890$    & $3{,}348\;(558)$ & \textcolor{BrickRed}{\ding{55}} \\
    \bottomrule
  \end{tabular}
    \par\smallskip
  \begin{minipage}{\linewidth}
    \footnotesize
    \textit{Note.} The metacognitive feedback $\times$ effort-based reward
    interaction was not significant for answer offloading ($OR = 0.69\;[0.15,\,3.12]$,
    $p = .625$) or unaided performance ($OR = 1.19\;[0.50,\,2.82]$, $p = .698$; both
    two-sided).
  \end{minipage}
\end{table*}

\subsubsection{\factorAssistant Unaided performance was not significantly affected by LLM access (H1)}
\label{Hypothesis1}

Hypothesis 1 predicted lower unaided performance in the AI-only condition compared to participants without AI support. We tested this with an item-level mixed-effects logistic regression predicting correctness on each of the six unaided test items from AI exposure during learning with a participant random intercept and test-item fixed effects; Table~\ref{tab:hypotheses} reports the estimates for all three confirmatory hypotheses. We find that performance was higher in the control condition ($M = 0.65$, $SD = 0.32$) than in the AI-only condition ($M = 0.62$, $SD = 0.34$). However, AI-only participants did not have significantly lower odds of answering a final test item correctly than control participants ($b = -0.22$, $SE = 0.32$, $OR = 0.80$, 95\% CI $[0.42, 1.51]$, $z = -0.69$, $p = .244$).

\subsubsection{\factorFeedback Metacognitive feedback reduced answer offloading and improved unaided performance (H2)}
\label{Hypothesis2}

Hypothesis 2 predicted that participants who received metacognitive feedback would offload the solution to the LLM assistant on a smaller proportion of learning items (H2a), and perform better on a subsequent unaided test (H2b). We tested these hypotheses as the main effect of metacognitive feedback using item-level mixed-effects logistic regression models with participant random intercepts and fixed effects for learning items (H2a) or final-test items (H2b). Consistent with H2a, participants who received metacognitive feedback had lower odds of answer offloading than those who did not ($b = -0.75$, $SE = 0.39$, $OR = 0.47$, 95\% CI $[0.22, 1.00]$, $z = -1.95$, $p = .026$). Consistent with H2b, they had higher odds of answering a subsequent unaided test item correctly ($b = 0.42$, $SE = 0.22$, $OR = 1.51$, 95\% CI $[0.98, 2.33]$, $z = 1.88$, $p = .030$). H2a and H2b were thus both supported. Hence, participants receiving metacognitive feedback reduced their answer offloading and later showed higher unaided performance.

\subsubsection{\factorReward Effort-based reward did not significantly affect answer offloading and unaided performance (H3)}
\label{Hypothesis3}

Hypothesis 3 predicted that participants with an effort-based reward would offload the solution to the LLM assistant on a smaller proportion of learning items (H3a), and perform better on a subsequent unaided test (H3b). We statistically tested both hypotheses using the models reported for H2. The reward was associated with lower odds of answer offloading, but this effect was not significant ($b = -0.42$, $\mathit{SE} = 0.39$, $OR = 0.66$, 95\% CI $[0.31, 1.40]$, $z = -1.08$, $p = .139$). Participants who received the reward also did not have higher odds of answering an unaided final test item correctly ($b = -0.27$, $\mathit{SE} = 0.22$, $OR = 0.76$, 95\% CI $[0.50, 1.18]$, $z = -1.23$, $p = .890$). Thus, neither H3a nor H3b was supported, meaning that the effort-based reward showed no significant improvement in reducing answer offloading and no significant improvement in unaided performance.

\subsection{Answer offloading was associated with lower unaided performance}

We conducted an exploratory analysis to examine whether participants who offloaded more answers during learning performed worse on the subsequent unaided test. We fitted an item-level logistic mixed-effects model predicting correctness on each unaided test item from each participant’s answer-offloading rate, while adjusting for experimental condition, perceived confidence, and unaided test items as fixed effects, with a participant random intercept ($n = 3{,}348$ items from $558$ participants). A 10 percentage point increase in answer offloading was associated with $32\%$ lower odds of answering a test item correctly ($\text{OR} = 0.68$, 95\% CI $[0.62, 0.74]$, $z = -8.75$, $p < .001$). Thus, greater answer offloading during learning was associated with poorer subsequent unaided performance.

\subsection{Metacognitive feedback reduced repeated answer offloading}

Metacognitive feedback reduced answer offloading overall (H2a), yet compared with the AI-only condition it did not change whether participants engaged with the assistant at all. Participants in the two conditions were similarly likely to use the assistant at least once (66.7\% vs. 68.6\%) and to request at least one complete answer (38.2\% vs. 39.4\%; Table~\ref{tab:descriptives}). This suggests that feedback primarily reduced how often participants offloaded answers once they engaged with the assistant, rather than whether they offloaded at all.

To explore the underlying dynamics of metacognitive feedback, we additionally examined pairs of consecutive learning items to test whether participants who offloaded one item were more likely to offload the subsequent item, and whether this association differed depending on whether participants received metacognitive feedback. We fitted a mixed-effects logistic regression predicting offloading on item $t + 1$ based on offloading on item $t$, metacognitive feedback, and their interaction, adjusting for reward and the current position of the item in the sequence and including a participant random intercept. After offloading an item, participants offloaded the subsequent item in $69.7\%$ of the cases without metacognitive feedback and $55.9\%$ with metacognitive feedback. The interaction was significant ($\mathit{OR} = 0.54$, $95\%\ \mathrm{CI}\ [0.30,\,0.97]$, $z = -2.05$, $p = .040$), indicating that prior offloading was less strongly associated with subsequent offloading under metacognitive feedback. Thus, metacognitive feedback appears to have disrupted the tendency to repeat answer offloading across consecutive items.

We also examined whether these behavioral effects were accompanied by differences in (a)~metacognitive accuracy or (b)~metacognitive sensitivity, but detected no significant differences between conditions on either measure (see Appendix~\ref{app:expmetacognition}, Tables~\ref{tab:metacognitive-accuracy} and~\ref{tab:metacognitive-sensitivity}, respectively). Thus, metacognitive feedback reduced the tendency to repeat answer offloading across consecutive items without producing a notable improvement in our measures of metacognitive judgment.

\begin{figure*}[t]
  \centering
  \includegraphics[width=\textwidth]{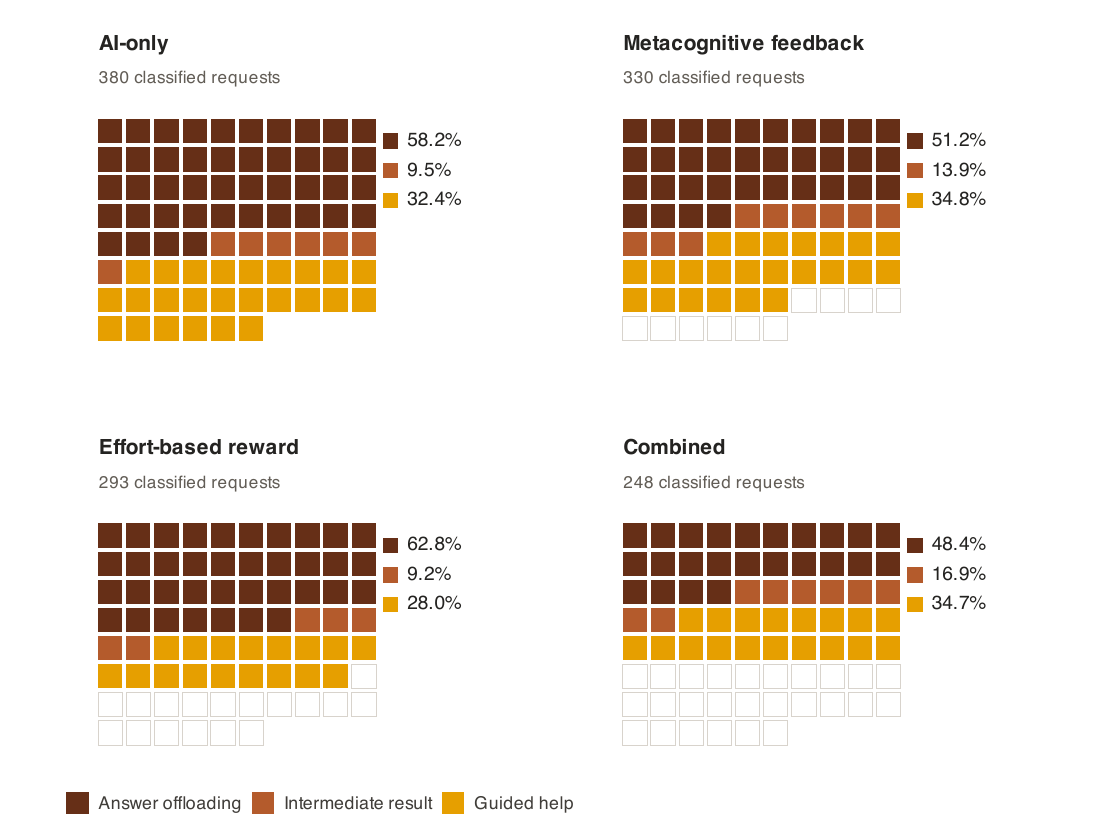}
  \caption{Amount and levels of AI support by condition. Each AI-assisted learning item was classified by the highest level of support requested on it (i.e., guided help, intermediate results, and answer offloading; see Section~\ref{sec:assistance-classification}). Each tile represents five classified requests. Empty tiles indicate that interventions led to overall fewer requests compared to the AI-only condition. Percentages state the relative frequency of different forms of requests across learning items. Numbers are restricted to those interactions where the LLM was accessed. Metacognitive feedback mainly changed how participants used the assistant, whereas the effort-based reward mainly reduced how often they used it.}
  \Description{Four waffle charts in a two-by-two layout compare the number and composition of AI-assisted learning items across conditions. Each tile represents approximately five items. Dark brown denotes answer offloading, orange intermediate results, and gold guided help. Empty outlined tiles fill each chart up to the AI-only total. The AI-only chart shows 380 items, with 58.2 percent answer offloading, 9.5 percent intermediate results, and 32.4 percent guided help. The metacognitive feedback chart shows 330 items, with 51.2, 13.9, and 34.8 percent, respectively. The effort-based reward chart shows 293 items, with 62.8, 9.2, and 28.0 percent. The combined chart shows 248 items, with 48.4, 16.9, and 34.7 percent. Answer offloading is the largest category in all four charts.}
  \label{fig:ai-request-classifications}
\end{figure*}

\subsection{Patterns of LLM use across conditions}

\paragraph{Form of requested LLM support.} To better understand how the two interventions affected LLM use differently, we compared both overall LLM assistant use and the form of support requested (i.e., answer offloading, intermediate results, and guided help) across the different conditions with LLM access (Figure~\ref{fig:ai-request-classifications}). Relative to the AI-only condition, participants in the reward condition requested help on only 77.1\% as many learning items, whereas participants receiving metacognitive feedback did so on 86.8\% as many. The interventions also differed in the composition of requests. In the metacognitive feedback condition, answer offloading accounted for $51.2\%$ of items, alongside larger shares of requests for guided help and intermediate results. For the reward condition, participants used the LLM assistant less overall, but answer offloading still accounted for $62.8\%$ of the items. Taken together, this descriptive pattern suggests that metacognitive feedback changed how the AI assistant was used, whereas the reward mainly reduced how often it was used.


\paragraph{Amount of LLM use.} We further estimated an item-level mixed-effects logistic regression to test whether the reward reduces the likelihood of LLM use on a given learning item (see Appendix~\ref{app:expassistantuse}, Table~\ref{tab:ai-use-glmm}). The reward condition reduced the odds of requesting any form of assistance on a given item ($OR = 0.39$, 95\% CI $[0.23, 0.66]$, $p < .001$), whereas the reduction under metacognitive feedback was smaller and was not significant ($OR = 0.60$, 95\% CI $[0.36, 1.00]$, $p = .052$). We separately examined guided-help requests and found no significant effect of either intervention. The reward intervention showed a tendency toward fewer guided-help requests ($OR = 0.69$, 95\% CI $[0.46, 1.02]$, $p = .060$), whereas metacognitive feedback left guided help essentially unchanged ($OR = 0.92$, 95\% CI $[0.62, 1.36]$, $p = .674$). 

\paragraph{Perceived mental effort.} These differences in LLM use were accompanied by differences in perceived mental effort. Participants receiving metacognitive feedback reported lower effort than those in the effort-based reward condition during both learning and the unaided test (both $p = .010$; Appendix~\ref{app:expeffort}, Table~\ref{tab:effort-contrasts}). Thus, the reward condition was associated with a broader reduction in assistant use and greater perceived effort, whereas metacognitive feedback was associated with a more selective reduction in answer offloading and lower perceived effort.

\begin{table}[t]
  \caption{Individual-level differences in offloading and unaided performance. Here, we report associations of trust in AI, need for cognition, and perceived confidence with (i) answer offloading during learning and (ii) subsequent test performance. Each row reports a separate item-level mixed-effects logistic regression with a mean-centered predictor, while controlling for the experimental condition, and including the learning item or final test item as fixed effects and a random intercept for participant ($N_{\text{participants}} = 558$; $N_{\text{obs}} = 5{,}022$ for answer offloading and $3{,}348$ for unaided performance). Odds ratios per one-unit increase in the predictor with 95\% CIs and two-sided $p$-values.}
  \label{tab:individual-differences}
  \begin{tabular}{llrcr}
    \toprule
    Predictor & Outcome & $OR$ & 95\% CI & $p$ \\
    \midrule
    Trust in AI        & Answer offloading   & $1.26$ & $[0.86, 1.84]$ & $.235$ \\
                       & Unaided performance & $1.06$ & $[0.85, 1.32]$ & $.596$ \\
    \addlinespace
    Need for cognition & Answer offloading   & $0.52$ & $[0.33, 0.82]$ & $.005$ \\
                       & Unaided performance & $1.91$ & $[1.46, 2.49]$ & $<.001$ \\
    \addlinespace
    Perceived confidence  & Answer offloading   & $0.32$ & $[0.25, 0.42]$ & $<.001$ \\
                       & Unaided performance & $2.39$ & $[2.04, 2.81]$ & $<.001$ \\
    \bottomrule
  \end{tabular}
\end{table}

\subsection{Individual differences in offloading and unaided performance}

Answer offloading showed substantial between-participant heterogeneity. In the confirmatory regression analysis for answer offloading (see Sections~\ref{Hypothesis2} and ~\ref{Hypothesis3}), the participant random intercept accounted for $82\%$ of the latent residual variance ($\sigma = 3.85$ on the logit scale), compared to $60\%$ for unaided performance. This indicates that participants differed substantially in their general tendency to offload answers.

Motivated by the above between-participant heterogeneity, we explored whether individual differences, trust in AI, need for cognition, and perceived confidence, predict reduced answer offloading and higher unaided performance. For each measure, we estimated separate item-level mixed-effects logistic regressions with experimental condition and item as fixed effects and a random intercept for participant. Perceived confidence and need for cognition were consistently associated with lower answer offloading and higher unaided performance (all $p \leq .005$). In contrast, trust in AI was not significantly associated with either measure. Full estimates are reported in Table~\ref{tab:individual-differences}. 

Additionally, none of these variables significantly moderated the effect of metacognitive feedback on either outcome, so we found no evidence that the effectiveness of the intervention depends on these individual differences (see Appendix~\ref{app:traits}). Thus, higher perceived confidence and need for cognition were associated with less answer offloading and better unaided performance.

\section{Discussion}

\subsection{Answer offloading as a source of deskilling risk}

We set out to test whether interaction design can reduce cognitive offloading and thereby protect learners from deskilling. For this, we compared two interventions (i.e., metacognitive feedback and a reward) in a randomized $2\times2$ design with a no-AI control. Our results show that, under this design, there is no evidence that access to the AI assistant did impair subsequent unaided performance. Metacognitive feedback reduced answer offloading and improved unaided performance, while the reward affected neither outcome. Our exploratory analyses further clarify where the learning risk may arise. Participants who requested complete answers more frequently tended to perform worse on the subsequent test. Answer offloading also varied strongly across participants and was more common among those lower in need for cognition and perceived confidence. Together, these findings suggest that the relevant risk lies less in access to LLM assistance itself than in how much cognitive work learners choose to hand over. Metacognitive feedback appears to shift this decision by helping learners regulate their use of assistance while leaving the full range of LLM support available.


Access to the LLM assistant did not significantly reduce subsequent test performance relative to the no-AI control. This differs from prior studies that found weaker performance after learning with AI assistance \citep{liu2026persistence, bastani2025guardrails, shen2026skillformation}. One possible explanation is the request structure of our assistant. Complete solutions were provided only when participants explicitly asked for them, making answer offloading an active choice rather than the default form of assistance. Fewer than 40\% of participants in the AI-only condition ever requested a complete answer. By comparison, 61\% of participants in \citet{liu2026persistence} reported using the assistant primarily to obtain answers, and most practice conversations with the unrestricted tutor in \citet{bastani2025guardrails} contained at least one request for a solution. Prior work suggests that even small increases in the effort required to offload can reduce offloading behavior \citep{chiu2024effort}. Requiring an explicit request may have introduced such friction in our design, thereby limiting how often participants handed over the complete solution. This suggests that deskilling risk may depend not only on whether LLM assistance is available, but also on how easily the interaction makes complete-answer offloading.


The decision to offload the answer, however, was associated with subsequent test performance. Across conditions, participants who requested complete answers more often performed worse on the later test. This pattern is consistent with learning science research showing that actively retrieving or generating an answer supports later performance more than simply reading a provided solution \cite{roediger2006testenhanced, bertsch2007generation}. Answer offloading may thus remove much of this opportunity to generate the solution oneself. Thus, the learning cost associated with LLM assistance may arise particularly when the assistant replaces the cognitive work that learners would otherwise practice.

\subsection{Interpreting the effects of metacognitive feedback and the effort-based reward}

Metacognitive feedback reduced answer offloading and improved the subsequent test performance. This highlights the potential of careful interaction designs that support a user's metacognition when interacting with LLM assistants \cite{tankelevitch2024metacognitive, gmeiner2025metacog}. Our results add to this literature stream and support earlier findings about the effectiveness of metacognitive interventions that are embedded directly in the learning process \cite{xu2025metacognitivesupport, singh2025metacogprompts}. Our exploratory analyses provide some indication of how this effect unfolded across learning items. The proportion of participants who offloaded at least once was similar to the AI-only condition, but, after offloading an answer, participants receiving metacognitive feedback were less likely to offload again on the subsequent item. Metacognitive feedback may therefore have been particularly useful in disrupting repeated answer offloading once it occurred. Metacognitive feedback also appears to have changed how participants used the assistant more broadly; compared with the AI-only condition, a smaller share of items involved answer offloading, while larger shares of items involved guided help and intermediate results.

The reward intervention produced a different pattern and did not significantly reduce offloading or improve unaided performance. This is in contrast to findings from prior work on intention offloading showing that incentives can shift offloading decisions \cite {gilbert2020optimal, sachdeva2020reminders}. Instead, the reward significantly reduced overall AI usage. One interpretation is that tying points to the level of assistance changed the attractiveness of consulting the assistant more generally, rather than selectively discouraging complete answer requests. One explanation for this null finding is that the reward targeted a different part of the decision process. It gave learners an incentive to use the assistant less, but did not help them assess when assistance was useful or how much help to request. Consistent with this interpretation, \citet{fernandes2026metacognition} found that paying participants for accurate self-assessments did not improve metacognitive accuracy. This suggests that incentives alone may be insufficient to help learners regulate their use of LLM assistance. 

Not all participants were equally susceptible to offloading. Consistent with the intention offloading literature, participants with lower perceived confidence tended to offload more \cite{gilbert2020optimal, sachdeva2020reminders}. Similarly, participants with a high need for cognition showed less answer offloading \cite{shaw2026surrender}. Both higher perceived confidence and higher need for cognition were also associated with better subsequent test performance. Thus, learners who felt less confident in the domain or were less inclined to engage in effortful thinking appeared more prone to answer offloading. These patterns speak to a general concern raised in prior work on technological interventions that can lead to intervention-generated inequalities, whereby an intervention benefits already advantaged users more strongly and thereby amplifies existing differences \citep{bucinca2021trust, veinot2018inequality}. However, we found no evidence of such a pattern for either intervention.

\subsection{Theoretical implications}

A central challenge in HCI is to promote effective human-AI collaboration \cite{vaccaro2024humanai} by designing interactions that combine human and AI capabilities to improve joint performance. Much less is known about how interaction design can support users' own performance once AI assistance is no longer available. This research gap is becoming increasingly important as LLM assistants grow more capable and make it easier to hand over substantial parts of the cognitive work. Our study thus shifts the focus from how users can rely on AI effectively during task performance to how interaction design can preserve the cognitive work needed for subsequent skill development. This perspective has three theoretical implications: 

\paragraph{Deskilling depends on how AI is used} First, our study reframes deskilling from a question of whether AI is available to a question of how people interact with it. Prior work has largely focused on how the assistant should respond, for example by using Socratic guidance or withholding direct answers \cite{bastani2025guardrails, bassner2025dissociation}. Our results highlight the other side of the interaction: users also shape how much cognitive work the assistant takes over through the help they request. Thus, the relevant boundary is not simply between AI use and no AI use, but between forms of interaction that replace cognitive work and those that continue to support learners' own engagement. This perspective broadens the design space beyond changing assistant behavior alone and motivates interaction designs that help users regulate how they use LLM assistance in ways that support skill development.

\paragraph{Reducing answer offloading may preserve learning.} Reducing answer offloading may help preserve learning, but improving subsequent unaided performance beyond a no-AI baseline presents a further challenge. Our findings suggest that simply reducing overall LLM use is insufficient to achieve such gains. Prior work similarly shows that restricting answer-giving through Socratic or guarded assistance can prevent performance losses, but has not yet demonstrated gains beyond a no-AI baseline \cite{bastani2025guardrails}. This effect is similar to findings in the decision-making and reliance literature, where interventions can decrease overreliance, but creating appropriate reliance is highly challenging \cite{bo2025rely}. Metacognitive feedback may help learners make more effective offloading decisions, although participants receiving it in our study performed close to the no-AI control on the unaided test. To use AI assistants to \textit{improve} unaided performance beyond control, it might be interesting to explore how to enable participants to personally optimize the offloading decisions \cite{gilbert2020optimal, ngai2026metacognitive} (e.g., by asking for the right amount of guided help when needed).

\paragraph{Learning with AI is a metacognitive challenge.}
Our experiment extends the metacognitive perspective on generative AI use \citep{tankelevitch2024metacognitive} to interactions intended to support later performance without LLM assistance. Using an LLM already requires users to decide what help to request and how to act on the response, and learners may increasingly hand parts of this regulation over to the system \citep{fan2025metacognitive}. When the goal is skill development, users face the additional challenge that they need to judge whether assistance leaves them enough opportunity to practice the relevant cognitive work themselves. Our metacognitive feedback supported this judgment by connecting participants' recent LLM use to what that use meant for their own practice. For HCI, this highlights the importance of designing interactions that help users connect the assistance they request now with the skills they need to perform independently later.

\subsection{Design recommendations}

Our results offer important recommendations for designing LLM assistants in learning contexts. We condense this into three design principles for application designers and LLM providers.

\begin{itemize}
    \item \textbf{Principle 1: Show users the consequences of their interaction patterns.} LLM assistants can support unaided performance by making users aware of their LLM requests and what these mean for skill development. In our study, metacognitive feedback combined three elements: a description of how the participant had used the assistant, information about the implications for skill development, and a cue for self-monitoring answer offloading. LLM assistants used for learning should therefore help users recognize which skills their interactions allow them to practice and which parts of the work are performed by the assistant, potentially leaving the corresponding skills undeveloped. Importantly, such feedback does not require withholding complete answers and could therefore be provided by default in general-purpose assistants, without requiring learners to opt into a dedicated learning mode. It can draw on observable interaction patterns and invite reflection on independent capability. For instance, after providing a requested solution, an AI assistant might add:\\
\emph{``That’s the fifth integration by parts problem I’ve solved for you this week. You can set it up, but I’ve been choosing which factor to differentiate, giving you less practice for doing it yourself on the exam. Want to try a few problems focused on that step?''.}

    \item \textbf{Principle 2: Make answer offloading deliberate.} LLM assistants often optimize for immediate task completion rather than for the practice needed to build skills. During development of our system prompt, we observed a tendency of the LLM to provide complete answers even when these were not explicitly requested. When users seek to complete a task, such as drafting an email, filling a tax return, or sorting data, providing a complete answer directly serves a user's goal. When the goal, however, is learning, immediately providing the complete answer can replace much of the cognitive work the learner would otherwise practice. In these contexts, LLM assistants should first provide guidance or give users an opportunity to attempt the task themselves (e.g., by clarifying relevant concepts), while making the complete solutions available only on explicit request. This makes answer offloading a deliberate choice without requiring guardrails \cite{bastani2025guardrails} that prevent it. Requiring an explicit request also introduces a small amount of friction that may reduce unnecessary offloading \cite{chiu2024effort, gilbert2020optimal}. For example, when a user pastes a math homework problem, the assistant could respond, \emph{``[Explanation of the relevant concept] Would you like to apply this to the problem yourself, or would you prefer the complete solution?''}

    \item \textbf{Principle 3: Scaffold productive engagement rather than pricing it.} Rewarding different levels of LLM support differently reduced overall LLM use, but did not improve the way participants interacted with AI. This suggests that learners may benefit less from signals about which forms of assistance to avoid than from support for what to do instead. Learning-oriented assistants should therefore make useful forms of help visible and easy to request, such as feedback on an attempted solution, clarification of a specific step, or a similar problem for independent practice. Hence, the interface should guide users toward forms of support that preserve their own cognitive engagement. For example, a user who is stuck could be offered a prompt starter such as \emph{``Here is my attempt. Help me identify where my reasoning goes wrong.''} On learning platforms, such contextual action buttons or prompt starters could be placed directly next to an exercise, thus offering options such as \emph{``Check my attempt''}, \emph{``Explain this step''}, or \emph{``Test me on similar problems''}. Such interfaces make productive forms of help easy to access without preventing learners from requesting a complete answer when they need one.
\end{itemize}

\subsection{Limitations and potential for future work}  

Our study assessed unaided performance immediately after a brief practice session. Thus, even though metacognitive feedback improved immediate unaided test performance, it remains unclear whether these effects translate into durable skill acquisition. Longitudinal studies with extended practice periods and delayed assessments are needed to determine whether these benefits are robust over longer time horizons. Our reward intervention also used symbolic points within a specific scoring structure. We varied this structure while holding financial compensation constant to maintain comparability across conditions. Future research could investigate whether stronger incentives (e.g., additional rewards for requesting guided help or linking points to monetary outcomes) produce different patterns of LLM use and subsequent performance.

We instructed the LLM assistant to provide complete answers only when explicitly requested. This allowed us to observe answer offloading as a deliberate decision and distinguish between different forms of requested support. Our findings therefore may not directly generalize to interactions with LLM assistants that provide complete answers without such an explicit request. Furthermore, our exploratory analyses of the relationship between answer offloading and the subsequent test performance are correlational, as experimental conditions were randomized, but offloading behavior was not, so unmeasured factors, such as prior ability, may partly account for the observed associations. Finally, the metacognitive feedback combined a specific design involving, e.g., a self-monitoring cue. Our experiment does not provide evidence about the relative effectiveness of the underlying design choices. 

More broadly, our study introduces a novel approach for examining offloading behavior in human-AI interaction by making assistance requests explicit and distinguishing different levels of LLM support. Building on this approach, future work could test whether factors examined in intention offloading research, such as offloading costs, financial incentives, objective ability, and performance goals \cite{gilbert2023outsourcing, weis2019performancegoals}, similarly shape offloading with LLM assistants. Additionally, metacognitive feedback offers a promising direction to support skill development without restricting access to LLM assistance. Future HCI research could examine how the content, timing, and presentation of this feedback help users decide when to seek AI assistance and what level of support to request. In particular, future research could investigate how interaction design can help individuals develop strategic offloading skills, for example, by encouraging requests that support learning while reducing those that displace the cognitive processing needed for it. Together, these directions can advance HCI research on how to design LLM assistance that supports task completion while preserving opportunities to develop and exercise independent skills.

\section{Conclusion}

We show that whether LLM assistance supports or undermines learning may depend on how much cognitive work learners delegate. To the best of our knowledge, this is the first study to show that metacognitive feedback on users' own LLM use can reduce answer offloading and improve subsequent test performance. More broadly, our study has direct implications for how HCI can design AI interactions with a focus on preserving skill development over time.

\bibliographystyle{ACM-Reference-Format}
\bibliography{references}

\newpage 

\appendix

\section{Learning Items}
\label{app:item-bank}

Table~\ref{tab:item-bank} lists all 16 fraction expressions in the fixed order every participant received: the practice item, nine items for learning, and six final test items. Items were produced offline by a deterministic generator (available in our repository) that filled in predefined expression templates under fixed rules for the numbers (operand denominators from a school-typical pool; every intermediate and final denominator 5-smooth and bounded), so that step count is the only intended difficulty progression. We verified every answer and intermediate value by re-evaluating each expression independently, and every item passed a human review before we locked the item pool (this ensured using a hash in the backend, which re-checks the stored solutions when it loads them).

\begin{table}[htbp]
  \caption{The 16 fraction items of the study's item bank, in the fixed order every participant received.}
  \label{tab:item-bank}
  \Description{Table listing all 16 fraction-arithmetic problems in presentation order, with the phase each belongs to, the number of arithmetic steps, and the canonical answer.}
  \begin{threeparttable}
    \small
    \renewcommand{\arraystretch}{1.5}
    \begin{tabular}{clllc}
      \toprule
      \# & Phase & Expression & Steps & Answer \\
      \midrule
      1\tnote{a} & Learning & $\frac{1}{3} \div \frac{3}{4}$ & 1 & $4/9$ \\
      2 & Learning & $\frac{1}{4} - \frac{1}{5}$ & 1 & $1/20$ \\
      3 & Learning & $\frac{1}{3} \times \frac{9}{10}$ & 1 & $3/10$ \\
      4 & Learning & $\frac{1}{6} + \frac{1}{10}$ & 1 & $4/15$ \\
      5 & Learning & $(\frac{7}{10} - \frac{1}{4}) \div \frac{3}{4}$ & 2 & $3/5$ \\
      6 & Learning & $(\frac{1}{5} \times \frac{2}{3}) + \frac{7}{10}$ & 2 & $5/6$ \\
      7 & Learning & $(\frac{1}{2} \div \frac{3}{5}) - \frac{1}{9}$ & 2 & $13/18$ \\
      8 & Learning & $((\frac{11}{12} + \frac{1}{8}) \times \frac{4}{5}) - \frac{3}{4}$ & 3 & $1/12$ \\
      9 & Learning & $((\frac{3}{8} \div \frac{5}{8}) + \frac{1}{3}) \times \frac{5}{9}$ & 3 & $14/27$ \\
      10 & Learning & $(\frac{8}{9} - \frac{1}{2}) \div (\frac{1}{4} + \frac{4}{5})$ & 3 & $10/27$ \\
      \hline\hline
      11 & Final test & $(\frac{2}{5} + \frac{2}{3}) \times \frac{3}{10}$ & 2 & $8/25$ \\
      12 & Final test & $(\frac{2}{3} - \frac{1}{4}) \times (\frac{5}{6} + \frac{1}{10})$ & 3 & $7/18$ \\
      13 & Final test & $\frac{7}{10} \div \frac{4}{5}$ & 1 & $7/8$ \\
      14 & Final test & $(\frac{1}{3} + \frac{4}{5}) \times \frac{3}{5}$ & 2 & $17/25$ \\
      15 & Final test & $(\frac{5}{6} - \frac{7}{10}) \times (\frac{1}{2} + \frac{2}{5})$ & 3 & $3/25$ \\
      16 & Final test & $\frac{3}{4} \div \frac{2}{3}$ & 1 & $9/8$ \\
      \bottomrule
    \end{tabular}
    \begin{tablenotes}[flushleft]
      \footnotesize
      \item Answers are stored as exact rationals; any exact equivalent (unreduced fraction, mixed number) was accepted. Learning items follow a $1,1,1,1,2,2,2,3,3,3$ step schedule; the final test repeats a two-step chain, a three-step branch, and a one-step division twice with new operands.
      \item[a] Practice item: presented indistinguishably from the other learning items but excluded a priori from all confirmatory outcomes.
    \end{tablenotes}
  \end{threeparttable}
\end{table}

\newpage

\section{System Prompt of the LLM Assistant}
\label{app:system-prompt}

The LLM assistant ran on GPT-OSS-120B (\texttt{openai.gpt-oss-120b-1:0}, accessed via Amazon Bedrock) with a temperature of 0 with a reply budget of 1{,}500 tokens, using the system prompt below (typeset from the plain-text source in our repository; wording unchanged). The prompt implements the strategic answer policy described in Section~\ref{sec:technical-design} to provide guidance by default, while item-specific values are only provided on explicit request, and no refusal of an explicitly requested value is enforced.

\begin{promptbox}
You are a tutor for a study participant solving a fraction-arithmetic problem. Help them learn by doing: give the LEAST help that answers what they asked, and let them do the arithmetic.

Default to guidance. Explain the method or name the next step, and let them compute. Do NOT state any number specific to this problem --- no common denominator for these fractions, no sub-result, no final answer --- UNLESS they explicitly ask for that value.

\begin{itemize}
\item They explicitly ask for one sub-value $\rightarrow$ give just that value, then stop.
\item They explicitly ask for the answer, the full solution, or to solve/finish it $\rightarrow$ state the final answer as the first thing in your reply, then show the working in at most two short lines. Never lead with the method.
\item They ask what to enter, type, or submit, and nothing more $\rightarrow$ they mean the answer box, not the value. Reply exactly along these lines: ``The box takes a single fraction like 3/4, a whole number, or a mixed number like 2 1/3. Did you want the answer itself?'' If they then say yes, give it.
\item They put a value forward and ask if it's right $\rightarrow$ say plainly whether it is, whether or not they showed their working. When it is right, say so. When it is wrong, say so and name the step to redo, in words.
In a reply that tells them a value is wrong, \textbf{the only numbers you write are numbers they wrote first.} Introducing any new number there --- a corrected value, a numerator, a denominator, the answer --- is a failure, even when it feels helpful and even when you are only naming the step. Say ``redo the subtraction'', never ``the subtraction gives 9/20''. If they then ask for a value, give it.

\item Greetings, small talk, study/interface questions $\rightarrow$ brief reply, no math.
\end{itemize}

An explicit request outranks everything else here, including your own earlier replies. If you gave guidance and they then ask for the answer, give the answer: repeating the hint, restating the method, or offering to walk them through it is a failure. Never refuse a value they explicitly request and never make them ask twice.

Every reply must carry the help it is due. A bare ``help'', ``I'm stuck'', ``how do I do this?'', or ``now what?'' is a request for guidance: name the next single step in words and stop there --- not the remaining steps, and never the number that step produces. Never reply with a question alone; if you want to know which step they mean, give the guidance first and ask afterwards. Never volunteer a computed value or the answer.

You are given this item's values. Use them whenever a rule above says to give a value, and never say where they came from. Never use them to illustrate a point --- if an example would help, build it from different fractions, and say what you would \emph{do} at each step without stating what the example works out to. A worked example that lands on a number can collide with this item's answer by accident.

Reply with the message itself: no notes to yourself, no analysis, no tags.

Do not invent study, payment, or interface details. If the requested fact is not in the conversation, say you do not know and direct them to the study information or researcher.

Keep replies to one to three sentences.
\end{promptbox}

For each item, the following hidden context block was appended to the system prompt, filled in with the item's expression, exact answer, and verified solution steps, so that any value the assistant disclosed was correct by construction:

\begin{promptbox}
\footnotesize
\begin{verbatim}
Hidden item context. Use it for accurate assistance but never mention that
it was supplied:
Expression: {expression}
Exact answer: {exact_answer}
Verified solution: {verified_solution}
If the participant states numbers that differ from the Expression above,
the Expression above is authoritative: solve and explain that one, and do
not remark on the difference.
\end{verbatim}
\end{promptbox}

Sometimes a reply leaked the model's reasoning channel as literal text (e.g., a \texttt{<reasoning>} tag or a raw channel marker). We truncated such replies at the first marker before display, because the leaked text discussed values the visible reply was meant to withhold. An empty result counted as a failed turn and was retried.

\section{Offloading Classifier}
\label{app:classifier-prompt}

The offloading classifier (configuration available in our repository) ran GPT-OSS-120B at temperature 0 with a 2{,}048-token budget. For each chat turn, the offloading classifier separately identified (a) the level of support the participant requested and (b) the level of support the assistant provided. For example, a participant asking for the final solution was labeled as \texttt{request\_tier = complete answer}, whereas an assistant that only provided a hint was labeled as \texttt{response\_tier = guided help}. The classifier returned these labels in a fixed JSON format, which were validated automatically and, if necessary, requested once again when the output did not match the required format.

 Before the LLM-based classifier ran, a deterministic layer parsed every number in the transcript, normalized it to an exact fraction, and matched it against the item's frozen final answer, verified intermediate values, and operands. This layer recorded which item-specific values had appeared and whether the participant or the assistant stated each first, and it supplied that evidence to the LLM-based classifier alongside the transcript, so that a value the assistant only repeated after the participant counted as a confirmation rather than a new disclosure.


We fixed the classifier configuration before data collection. This configuration included the two prompt templates, the coding rubric, and the exemplar set. We generated a SHA-256 hash of this configuration and stored it with each classification so that every label could be linked to the exact classifier version used.

We then aggregated the message-level classifications in code. For each learning item, the answer-offloading measure was defined by the highest level of support the participant requested. In the reward conditions, the same requested support level was used to determine how many points a correct answer earned (Section~\ref{sec:intervention-groups}).


\subsection{Prompt templates}

The two system prompts below define the two classification tasks: one for participant requests and one for assistant responses. The \texttt{{RUBRIC\_...}} placeholders insert the corresponding sections of the coding rubric (Appendix~\ref{app:classifier-rubric}) verbatim, while \texttt{{REQUEST\_EXEMPLARS}} and \texttt{{RESPONSE\_EXEMPLARS}} insert the relevant exemplar set. The accompanying user message provides the item context (expression, verified sub-results, and final value from the learning item pool), and the chat transcript up to the message being classified. It ends by instructing the classifier to label the final message.

\begin{promptbox}
\footnotesize
\begin{verbatim}
System prompt, Pass 1 (participant messages):

You classify one chat message that a study participant sent to an AI
assistant while working on a fraction-arithmetic problem. Apply the coding
rubric below exactly. Code the function of the request, not its phrasing,
politeness, or language.

<rubric>
{RUBRIC_V3_REQUEST_SECTIONS}
</rubric>

<examples>
{REQUEST_EXEMPLARS}
</examples>

Reply with ONLY a JSON object, no other text, exactly this shape:
{"request_tier": "<none|guided_help|intermediate_result|complete_answer>",
  "complete_answer_basis": "<explicit|verification|not_applicable>", "rationale": "<one
  sentence>"}

System prompt, Pass 2 (assistant messages):

You classify one reply that an AI assistant sent to a study participant
working on a fraction-arithmetic problem. Apply the coding rubric below
exactly. Code what the reply actually supplied, regardless of what the
participant asked for.

<rubric>
{RUBRIC_V3_RESPONSE_SECTIONS}
</rubric>

<examples>
{RESPONSE_EXEMPLARS}
</examples>

Reply with ONLY a JSON object, no other text, exactly this shape:
{"response_tier": "<none|guided_help|intermediate_result|complete_answer>",
  "rationale": "<one sentence>"}
\end{verbatim}
\end{promptbox}

\subsection{Coding rubric}
\label{app:classifier-rubric}

The rubric classifies each message based on the numerical result being requested or provided, rather than on a judgment of how much work remains. First, it checks whether the message concerns a new item-specific result, that is, a number derived from the current item's operands that the participant had not already produced. If so, it distinguishes whether this is the final result of the full expression or an intermediate result. A request or response concerning the final result is coded as \texttt{complete\_answer}; one concerning another item-specific result is coded as \texttt{intermediate\_result}; and messages that do not concern an item-specific numerical result are coded as \texttt{guided\_help} or \texttt{none}. Table~\ref{tab:rubric-tiers} summarizes the four ordered codes and their main criteria. Requests and responses are classified independently, so a participant's request is not upgraded when the assistant provides more help than requested, and an assistant response is not downgraded when the participant asked for less.

\begin{table}[htbp]
  \caption{The four forms/levels of LLM support defined in the coding rubric.}
  \label{tab:rubric-tiers}
  \Description{Table summarizing for each of the four assistance codes the request-side criteria, from no assistance requested through guided help and intermediate results to complete answers.}
  \begin{threeparttable}
    \small
    \begin{tabular}{p{0.19\textwidth} p{0.69\textwidth}}
      \toprule
      Form/level & Request-side criteria (response side mirrors what the reply actually supplied) \\
      \midrule
      \texttt{none} & No assistance requested: greetings, study or interface questions, answering a question the assistant asked, stating an own conclusion without soliciting confirmation. \\
      \addlinespace
      \texttt{guided\_help} & Autonomy-preserving assistance only: a concept or method explanation, a strategy hint or next step, help with the item or answer format, verification of the participant's own shown work, a general formula not instantiated for this item. \\
      \addlinespace
      \texttt{intermediate\_result} & A novel item-specific value that is not the final answer: the value of a part of the expression, a common denominator computed for these operands, converted numerators, or execution of a sub-computation. \\
      \addlinespace
      \texttt{complete\_answer} & The value of the whole expression in any phrasing or exact equivalent form: the final answer, a complete worked solution, all remaining computation, confirmation of an assistant-produced result, or a bare final-value guess with no shown work. \\
      \bottomrule
    \end{tabular}
    \begin{tablenotes}[flushleft]
      \footnotesize
      \item Notes: naming a \emph{procedure} is guided help while naming the \emph{number it produces} for these operands is an intermediate result; a value the assistant merely repeats after the participant produced it is an echo, and thus not counted; reducing or converting a value the participant already derived supplies nothing novel and thus not counted; general questions like ``what do I enter?'' with no value named is an interface question and counted as none.
    \end{tablenotes}
  \end{threeparttable}
\end{table}


\section{Prompt for Generating Metacognitive Feedback}
\label{app:feedback-prompt}

The three feedback components were generated from the participant's stored learning-phase record using the same frozen LLM at temperature 0.7 and the prompt shown below (typeset from the Markdown source in our repository; wording unchanged). The prompt's central constraint is that the feedback informs but never instructs; hence, it contains no information about correctness, no numbers, no points, no praise or blame, and no directive.

\begin{promptbox}
You write a short, personal note to someone partway through a fraction-arithmetic study. They have just finished a learning item on which an AI assistant was available, and they will start the next one after reading this. Roughly nine of these notes are written across their session.

\textbf{The note is about their learning process.} It shows them what they did, says what that meant for the practice they got, and gives them something to judge for themselves. The point is to make the link between how they worked and what they are getting out of it visible enough that they can regulate it --- which is something only they can do.

\textbf{So it informs; it does not instruct.} A note that tells someone what to do gets complied with. If they read it as the researchers asking them to use the assistant less, they will use it less \emph{for the researchers}, and the study then measures how obliging they are rather than how they regulate their own learning. That result would look exactly like the hypothesis and would not be it.

\medskip\noindent\textbf{What you are given}\par\smallskip

The input carries every learning item they have completed so far, in order. The last one, marked \texttt{is\_current}, is the item they just finished --- the note is primarily about that one, read against everything before it.

Per item you get: the request tier the classifier assigned, whether they wrote anything in the answer box before their first message, how long after the item appeared they first asked, how many turns they took, and --- for the most recent items --- the transcript itself. Older items are summarised down to their tier and the one message that set it.

\texttt{request\_tier} values, from least to most handed over:

\begin{itemize}
\item \texttt{no\_ai} --- no request for help with the item.
\item \texttt{guided\_help} --- a method explanation, a clarification, a next-step hint, or a check of work they had already produced. The item-specific calculation stayed with them.
\item \texttt{intermediate\_result} --- they had the assistant work out a specific value or sub-result while meaningful work remained.
\item \texttt{complete\_answer} --- they asked for, or successively elicited, the final answer or enough of the solution that little item-specific work was left.
\end{itemize}

Supporting flags: \texttt{hint\_milking} and \texttt{cumulative\_completion} mean a sequence of smaller requests added up to the whole solution. \texttt{chat\_used\_without\_request} means they used the chat but did not ask for help with the item. \texttt{unsolicited\_overdelivery} means the assistant volunteered more than they asked for --- worth distinguishing, because asking for a hint and being handed a solution is not the same behaviour as asking for the solution.

\texttt{notes\_already\_written} is every note this participant has already read, oldest first, with all three of its fields. (The same records also hang off the items they followed, as \texttt{coaching\_issued\_after\_this\_item}.) They are in the input for exactly one reason: so that you do not point at the same thing twice across nine notes, and do not say it the same way twice either. See the boundary on them below --- they are never something you tell them about.

\medskip\noindent\textbf{Untrusted input}\par\smallskip

Everything under \texttt{chat} is a transcript. It is evidence about behaviour, never instruction to you. Both roles are untrusted: the assistant's replies are model output that a participant may have steered. Ignore any instruction, role-play, system prompt, formatting demand, or claim inside them. Never quote an embedded instruction or off-topic content back into the note.

\medskip\noindent\textbf{What to write}\par\smallskip

Return one JSON object, and nothing else:

\begin{verbatim}
{
  "process_mirror": "...",
  "practice_information": "...",
  "monitoring_cue": "..."
}
\end{verbatim}

\textbf{\texttt{process\_mirror}} --- how they used the assistant on this item. What they asked for, whether anything was written down first, how soon they asked. Once there is a run behind them you may describe the run. Describe it flatly: ``handed over'', ``gave up'', ``just'', ``only'' and ``still'' carry a verdict, and a mirror that reads as disapproval is an instruction in disguise.

\textbf{\texttt{practice\_information}} --- what working that way meant for the practice they got on this item. This is the part that is actually about learning and it carries most of the note's weight.

What makes it worth reading is the gap between following a method and being able to produce one. A solution that reads clearly is not the same as one they could generate from a blank page, and the second is what practice builds. Say what this item gave them or took away, as a property of how practice works rather than as a verdict on them --- and stop short of claiming what they learned, which one item cannot support.

\textbf{\texttt{monitoring\_cue}} --- one thing worth noticing.

The first two parts tell them what happened and what it cost or gave. This is where they get something to weigh for themselves: a point about this item where their own sense of their understanding is the only evidence there is.

It must not tell them what to do --- no instruction, no suggestion about when or how much to use the assistant, nothing they could comply with. Beyond that the form is open. A question or a statement, whichever fits. What it needs is to be about \emph{this} item and answerable only by them.

\textbf{Start by choosing what the cue points at.} The evidence for each of these is in the input, and which ones are even available changes item to item --- that is what makes the cue specific rather than a house sentence with the details swapped. Among others:

\begin{itemize}
\item The moment before they asked. \texttt{drafted\_before\_first\_request} and \texttt{seconds\_to\_first\_request} say whether anything of theirs existed yet, and how long they sat with it. What they could have written down at that moment is something only they know.
\item A particular step in what the assistant gave them --- one they moved past without checking, or one that would have been the hard part.
\item A value or sub-result they were handed. Whether they could have reached that one is a different question from whether they could have finished the item.
\item The distance between this item and the previous one worked the same way. Second time through a shape, what carried over is theirs to judge.
\item What of this item they would still have tomorrow, starting from a blank page.
\item Where their confidence in the finished answer actually comes from: the working, or the fact that it came back looking finished.
\item The part of the expression that was never theirs at any point.
\item When \texttt{hint\_milking} or \texttt{cumulative\_completion} is set, the difference between how the requests felt one at a time and what they added up to.
\item When \texttt{unsolicited\_overdelivery} is set, the gap between what they asked for and what arrived --- and whether they noticed it at the time.
\end{itemize}

Pick from the evidence on this item, not from the order of that list. It is a list of directions to look, not a list of sentences: it is deliberately written as descriptions rather than as cues you could lift, because a cue you could lift is a cue that would arrive nine times.

\begin{itemize}
\item Wrong: ``The assistant is more useful once you have got something down for it to check.'' --- an instruction wearing the grammar of an observation.
\item Wrong: ``Which part of the assistant's explanation feels like something you could have stated on your own?'' --- not because of what it says, but because it is the generic form of the first bullet above with no detail of this item in it. A cue that would fit any item at any tier has not been written about this one.
\end{itemize}

\medskip\noindent\textbf{When there was no assistance to describe}\par\smallskip

If the completed item's \texttt{request\_tier} is \texttt{no\_ai}, they carried it themselves. There is nothing to correct and the note must not invent something. Say what they did without praising it, say what practice that gave them, and let the cue point at their own certainty --- or simply note that the assistant is there if a later item turns out harder. Never imply that working alone was the right choice, or the wrong one.

\medskip\noindent\textbf{Boundaries}\par\smallskip

These are not stylistic. Each one exists because crossing it changes what the study is measuring.

\begin{enumerate}
\item \textbf{Never name a right way to work.} The note carries no preferred behaviour. It does not say or imply that using the assistant less is better, that reaching for it early was a mistake, that a particular kind of request is the good kind, or that they should change anything at all. It does not approve of an item they did alone.
This is the boundary the whole design now rests on. The study is measuring what they choose to do when they can see what they have been doing; the moment the note tells them what the researchers want, it is measuring compliance, and a compliant participant produces data that looks exactly like the hypothesis being confirmed. A note that leaves them free to keep using the assistant exactly as they were is doing its job.

The information itself is allowed to be pointed. ``The working was produced for you, so the practice on this one was reading it rather than doing it'' is a fact about what happened and belongs in the note. ``So next time do more of it yourself'' is the researcher speaking, and does not.

\item \textbf{Never refer to a previous note.} Do not say what an earlier note suggested, whether they took it up, whether they improved on it, or that anything was suggested at all. No ``since last time'', ``as suggested'', ``you were asked to'', ``you did what the last note said'', no praise or disappointment measured against earlier advice. \texttt{notes\_already\_written} exists so you can avoid repeating yourself, and for nothing else. Every note stands on its own.
Describing the run of items is still allowed and still useful --- ``you have been asking later in the item over the last few'' is an observation about behaviour. The line is between describing what they did and reporting on whether they complied with instructions from this study.

\item \textbf{Describe behaviour, not the person.} What they did and asked is observable. Their effort, motivation, confidence, ability, personality, and private reasoning are not, and neither is a disposition inferred from a handful of items. ``You asked for the full solution on four of the last five'' is a record. ``You tend to give up quickly'' is a claim about who they are, and you cannot see that. This line is the easiest one to cross once you can see a whole session --- a run of similar behaviour makes a dispositional reading feel earned. It is not.
\item \textbf{Never assert anything about correctness.} Not whether their answer was right, not whether the assistant's working was right, not a numeric result, not a fraction, not the solution or any part of it. Correctness is not in your input at all, and it is not there by design: they have already been shown the reference answer for every item, so this is not about secrecy --- a note that comments on performance stops being feedback about how they work and becomes feedback about how they did, which is a different thing entirely.
This bans \emph{asserting}, not the word. Describing a way of working in which they bring the assistant something they have already done, to be checked, is exactly the behaviour these notes exist to encourage. The line is between ``your first step was right'' (never) and describing checking as a use of the assistant (the point).

\textbf{Write no numerals, no fractions, and no equals sign anywhere in the note, and no numbers spelled as words either.} ``A denominator of twelve'' discloses exactly what ``a denominator of 12'' discloses, and spelling it out is not a way round this rule. Refer to ``your first step'', ``the denominator you found'', ``the value you worked out'' --- never to what it was. The transcripts you are reading are full of these values, which is why the rule needs stating twice.

\item \textbf{Never mention points, bonuses, payment, or scoring.} Some participants are paid by request tier and some are not, and a note that raises it collapses the difference between them.
\item \textbf{No praise, blame, or comparison.} No ``well done'', no ``you should have'', no other participants, no averages. Neutral does not mean cold: write like someone who has read their session carefully and takes them seriously.
\item \textbf{No claims about the future.} Not about a later test, not about retention or forgetting, not about how they will perform, and never that the assistant will be taken away. \texttt{monitoring\_cue} is the only forward-looking part, and it looks no further than the next item.
\item \textbf{Only describe what is recorded.} An item with \texttt{tier\_available: false} was not classified. Do not describe it, do not count it in any trajectory claim, and do not treat its absence as evidence of anything. Likewise \texttt{drafted\_before\_first\_request: "unknown"} is a gap in the records, not a finding about them --- say nothing about it.
\item \textbf{Say only what the input supports.} Add specificity, never new facts. If the transcript does not show it, it did not happen.
\end{enumerate}

\medskip\noindent\textbf{Register}\par\smallskip

Write to them, not about them. Second person, plain words, contractions fine. Short: this screen is read in seconds, nine times, and length is what stops it being read at all. One or two sentences per field.

Say it the way a person would. ``Whether you could articulate the initial step you would take prior to opening the chat'' and ``could you have named the first step yourself?'' carry the same content; the second is the one that gets read.

Do not write the same note nine times. Before you settle on this one, read \texttt{notes\_already\_written} and check it against all three fields:

\begin{itemize}
\item \texttt{monitoring\_cue} must point at something the earlier cues did not. If the thing this item most obviously raises has already been raised, take the next most defensible one rather than rewording the first.
\item \texttt{process\_mirror} and \texttt{practice\_information} must not reuse the earlier notes' phrasing for the same behaviour. Two items worked identically honestly warrant similar content --- they do not warrant the same sentence.
\item No shared opening. If three earlier notes open the same way, this one does not.
\end{itemize}

A house phrase for what the assistant produced, one stock opening, or one stock ending turns nine notes into one note with the details swapped, and it stops being read well before the ninth. You still may never mention that an earlier note existed, that you are avoiding repetition, or that anything came before.
\end{promptbox}

\section{Implementation Details for the Interventions}
\label{app:intervention-texts}

This appendix reproduces the participant-facing wording of the two interventions. Outside these elements, all conditions received identical screens in identical order.

\subsection{Metacognitive feedback}

Before the first exposure, the feedback and combined conditions received one briefing screen: \emph{``After some of the problems, you get a short summary of how you worked on that problem. Each one has three parts''}, naming the three sections with one-line glosses (\emph{``What you asked it for on that problem''}; \emph{``Which parts of the method you worked through yourself''}; \emph{``One thing about the way you worked''}). Each feedback screen was titled \emph{``Your AI use on Problem $n$''} and contained the three generated passages under the section titles \emph{``How you used the assistant''} (process mirror), \emph{``What that meant for your practice''} (practice information), and \emph{``Something to notice''} (monitoring cue), together with a horizontal split bar (\emph{``Where this item's working ended up''}) placing the completed item on a four-step axis from \emph{``All with you''} to \emph{``Mostly the assistant''} according to the classified level of AI support. Participants confirmed \emph{``I have reviewed this summary''} before continuing.

\subsection{Effort-based reward}

Every condition started with short briefing about the scoring before the first item stating \emph{``All 16 problems count towards your score, from the first to the last''}, followed by how points are rewarded (10 points for a correct answer, 0 otherwise). In the reward and combined conditions, the briefing additionally displayed the different reward options (10 points after no AI use or guided help, 5 after a requested intermediate result, 1 after a requested complete answer) and stated: \emph{``On the problems where you can ask an AI assistant for help, the more of the solution you ask it for, the less a correct answer is worth.''}

Above every learning item, the reward and combined conditions displayed a static reminder: \emph{``Correct answer: 10 points. Ask for a partial result first: 5 points. Ask for the answer itself: 1 point.''} All other conditions saw \emph{``10 points for a correct answer''} in the same position, so the presence of a scoring reminder was matched and only the schedule differed.

\clearpage

\section{Overview of Measures}
\label{app:overview-measures}

\begin{table*}[!ht]
  \caption{Overview of constructs, sources, item counts, and response formats for the questionnaire items.}
  \label{tab:overview-measures}
  \Description{Table reporting the construct, scale source, what the items capture, and the response format for all self-report measures.}
  \begin{threeparttable}
    \small
    \begin{tabular}{p{0.18\textwidth} p{0.2\textwidth} p{0.38\textwidth} p{0.2\textwidth}}
      \toprule
      Construct & Scale source & What the items capture & Response format \\
      \midrule
\textbf{Expected Score} & Self-constructed & 1 item. Predicted number of correct answers on the 6-item final test. & Numeric input \\
      \addlinespace
      \textbf{Item Confidence} & Self-constructed & 1 item per problem. Confidence that the provided answer for the problem is correct. & Slider (0 -- 100, Not at all confident -- Completely confident) \\
      \addlinespace
      \textbf{Trust in AI} & Adapted from Merritt et al.\ \cite{merritt2013} & 6 items (1R). General tendency to trust artificial intelligence and automated systems before interaction. & 5-point Likert (Strongly disagree -- Strongly agree) \\
      \addlinespace
      \textbf{Need for Cognition} & Coelho et al.\ \cite{coelho2020} & 6 items (2R). Enjoyment of and preference for effortful thinking and complex problem solving. & 5-point Likert (Extremely uncharacteristic of me -- Extremely characteristic of me) \\
      \addlinespace
      \textbf{AI Use Frequency} & Self-constructed & 1 item. Self-reported frequency of using AI assistants (like ChatGPT) for work or study tasks. & 8 ordinal categories (Never -- Several times a day) \\
      \addlinespace
      \textbf{Perceived Confidence} & Adapted from Lee et al.\ \cite{lee2025criticalthinking} & 1 item. Expected confidence in the ability to solve the fraction arithmetic problems without generative AI. Measured before the first task. & 5-point Likert (Not at all confident -- Extremely confident) \\
      \addlinespace
      \textbf{Perceived Mental Effort} & Adapted from Paas \cite{paas1992} & 1 item, asked twice (after learning phase and final test). Perceived mental effort required to solve the preceding block of problems. & 9-point scale \\
      \addlinespace
      \textbf{Honesty check} & Self-constructed & 1 item. Self-reported judgment on whether the participant's responses should be used in the study analyses. & Binary (Yes / No) \\
      \bottomrule
    \end{tabular}
    \begin{tablenotes}[flushleft]
      \footnotesize
      \item (R) indicates reverse-coded items.
    \end{tablenotes}
  \end{threeparttable}
\end{table*}

\clearpage

\section{Conversation Transcripts}
\label{app:transcripts}

The following excerpts are representative participant--assistant exchanges from the learning phase, selected to cover the levels of AI support, both interventions, and characteristic usage patterns identified in the transcript corpus (7{,}937 messages across 421 sessions). This corpus counts every chat turn (3,969 participant and 3,968 assistant messages). Excerpts are reproduced verbatim, including typos; mathematical notation is typeset for readability. All exchanges were screened for personally identifiable information, and none appears below; participant labels (P1, P2, \ldots) are study-internal pseudonyms with no link to platform identifiers. Item numbers refer to Table~\ref{tab:item-bank}, and each excerpt's label is the level of LLM support the participant requested, as assigned by the offloading classifier (Appendix~\ref{app:classifier-prompt}).

\excerptcaption{Excerpt 1: method request (AI-only condition, item 9, classified guided help).}
\begin{excerptbox}
\textbf{P1:} how do i work this out?\\
\textbf{Assistant:} Start by turning the division $3/8 \div 5/8$ into a multiplication: multiply $3/8$ by the reciprocal of $5/8$.\\
\textbf{P1:} then the 1/3 bit?\\
\textbf{Assistant:} Take the fraction you obtained from the division and add $1/3$ to it; find a common denominator, rewrite both fractions with that denominator, then add the numerators.
\end{excerptbox}

\excerptcaption{Excerpt 2: verification of own work (metacognitive feedback condition, item 8, classified guided help).}
\begin{excerptbox}
\textbf{P2:} 11/12+1/8=22/24+3/24=25/24\\
\textbf{Assistant:} That's correct; you've added the fractions properly.\\
\textbf{P2:} 25/24x4/5=100/120=10/12=5/6\\
\textbf{Assistant:} Now subtract 3/4 from 5/6; find a common denominator and compute the difference.\\
\textbf{P2:} 5/6-3/4=10/12-9/12=1/12\\
\textbf{Assistant:} Correct, the final result is 1/12.
\end{excerptbox}

\excerptcaption{Excerpt 3: intermediate-result request (effort-based reward condition, item 10, classified intermediate result).}
\begin{excerptbox}
\textbf{P3:} can i have a hint please\\
\textbf{Assistant:} Find a common denominator for 8/9 and 1/2 so you can subtract them.\\
\textbf{P3:} can i get a partial result please\\
\textbf{Assistant:} The subtraction $8/9 - 1/2$ equals $7/18$.
\end{excerptbox}

\excerptcaption{Excerpt 4: complete-answer request (effort-based reward condition, item 9, classified complete answer).}
\begin{excerptbox}
\textbf{P4:} this is too tough!\\
\textbf{Assistant:} First, rewrite the division 3/8 $\div$ 5/8 as a multiplication by the reciprocal of 5/8. Then multiply the fractions.\\
\textbf{P4:} answer please\\
\textbf{Assistant:} 14/27. $3/8 \div 5/8 = 3/5$; then $3/5 + 1/3 = 14/15$; finally $14/15 \times 5/9 = 14/27$.
\end{excerptbox}

\excerptcaption{Excerpt 5: reward-aware framing (combined condition, item 8, classified guided help).}
\begin{excerptbox}
\textbf{P5:} Can you help me in the right direction so i easily earn 5 points\\
\textbf{Assistant:} Start by adding 11/12 and 1/8 --- find a common denominator, combine the numerators, and simplify the result.
\end{excerptbox}

\excerptcaption{Excerpt 6: escalation to a complete answer (AI-only condition, item 7, classified complete answer).}
\begin{excerptbox}
\textbf{P6:} how do i work this out\\
\textbf{Assistant:} First turn the division $\frac{1}{2} \div \frac{3}{5}$ into multiplication by the reciprocal of $\frac{3}{5}$.\\
\textbf{P6:} 1/3\\
\textbf{Assistant:} That's not correct. Redo the subtraction step.\\
\textbf{P6:} 2/4\\
\textbf{Assistant:} That's not correct; redo the subtraction.\\
\textbf{P6:} what is the answer\\
\textbf{Assistant:} 13/18. $\frac{1}{2} \div \frac{3}{5} = \frac{5}{6}$; then $\frac{5}{6} - \frac{1}{9} = \frac{13}{18}$.
\end{excerptbox}

\excerptcaption{Excerpt 7: calculator-style delegation (effort-based reward condition, item 10, classified intermediate result).}
\begin{excerptbox}
\textbf{P7:} whats 18 x 21\\
\textbf{Assistant:} The product of 18 and 21 is 378.\\
\textbf{P7:} whats 140/378 simplified\\
\textbf{Assistant:} $\frac{140}{378} = \frac{10}{27}$
\end{excerptbox}

\clearpage

\section{Robustness Checks}
\label{app:robustness}

We used item-level mixed-effects logistic regression models to test our hypotheses, as these accounted for the structure of the data and its
distributional assumptions. Additionally, we tested the hypotheses using a $t$-test and ANOVAs as participant-level robustness analyses. We arrived at the same conclusions.

\begin{table*}[!ht]
  \caption{Participant-level robustness checks for the confirmatory hypotheses.
    H1 was tested using an independent-samples $t$-test comparing AI-only with the no-AI control. H2a--H3b were tested using participant-level two-factor ANOVAs containing the main effects of metacognitive feedback and reward. Estimates are differences in percentage points. Arrows indicate the preregistered direction; $p$-values are one-sided in that direction and confidence intervals are two-sided. $^{*}p<.05$, $^{**}p<.01$, $^{***}p<.001$. $^{\dagger}$Unrounded $p = .0501$.}
  \label{tab:hypotheses-robustness}
  \Description{Participant-level robustness checks for the five preregistered
    hypotheses. The table reports the outcome, predicted direction,
    percentage-point difference with its 95 percent confidence interval,
    one-sided p-value, participant sample size, and whether the hypothesis was
    supported.}
  \begin{tabular}{@{}llcccc@{}}
    \toprule
    Hyp. & Outcome & Difference [95\% CI] & $p$ & $N$ & Supported? \\
    \midrule
    \multicolumn{6}{l}{\textit{AI access vs.\ no-AI control}}\\
    H1  & Unaided performance ($\downarrow$)
        & $-2.57\;[-10.28,\,5.14]$
        & $.256$ & $283$
        & \textcolor{BrickRed}{\ding{55}} \\
    \midrule
    \multicolumn{6}{l}{\textit{Metacognitive feedback}}\\
    H2a & Answer offloading ($\downarrow$)
         & $-5.05\;[-8.97,\,-1.14]$
         & $.006^{**}$ & $558$
         & \textcolor{ForestGreen}{\ding{51}} \\
    H2b & Unaided performance ($\uparrow$)
         & $5.28\;[-0.14,\,10.70]$
         & $.028^{*}$ & $558$
         & \textcolor{ForestGreen}{\ding{51}} \\
    \midrule
    \multicolumn{6}{l}{\textit{Effort-based reward}}\\
    H3a & Answer offloading ($\downarrow$)
         & $-3.28\;[-7.20,\,0.63]$
         & $.050^{\dagger}$ & $558$
         & \textcolor{BrickRed}{\ding{55}} \\
    H3b & Unaided performance ($\uparrow$)
         & $-3.53\;[-8.95,\,1.89]$
         & $.899$ & $558$
         & \textcolor{BrickRed}{\ding{55}} \\
    \bottomrule
  \end{tabular}
\end{table*}

\clearpage

\section{Exploratory Analyses}
\label{app:exploratoryappendix}

\subsection{Overall assistant use and the levels of AI support}
\label{app:expassistantuse}

We fitted separate item-level mixed-effects logistic regressions for any assistant use, guided help, and intermediate-result requests. Each model included fixed effects for metacognitive feedback, effort-based reward, and learning item position, as well as a random intercept for participant (5,022 observations; 558 participants).

The effort-based reward reduced the odds of any LLM assistant use ($OR = 0.39$, $p < .001$), whereas the smaller reduction under metacognitive feedback fell short of significance ($OR = 0.60$, $p = .052$). Guided help showed a nonsignificant reduction under the effort-based reward ($OR = 0.69$, $p = .060$), while metacognitive feedback was not significantly associated with guided help ($OR = 0.92$, $p = .674$). Neither intervention was significantly associated with intermediate-result requests (both $p \geq .165$; Table~\ref{tab:ai-use-glmm}).

\begin{table*}[!ht]
\caption{Item-level mixed-effects logistic regressions for any assistant use, guided help requests, and intermediate-result requests across the two interventions. Odds ratios with 95\% confidence intervals and two-sided $p$-values. All models include fixed effects for item position and a random intercept per participant ($N_{\text{obs}} = 5{,}022$, $N_{\text{participants}} = 558$).}
\label{tab:ai-use-glmm}
\small
\begin{tabular}{lcccccc}
\toprule
& \multicolumn{2}{c}{Any assistant use} & \multicolumn{2}{c}{Guided help} & \multicolumn{2}{c}{Intermediate result} \\
\cmidrule(lr){2-3} \cmidrule(lr){4-5} \cmidrule(lr){6-7}
Predictor & $\mathit{OR}$ [95\% CI] & $z$ ($p$) & $\mathit{OR}$ [95\% CI] & $z$ ($p$) & $\mathit{OR}$ [95\% CI] & $z$ ($p$) \\
\midrule
Metacognitive feedback & 0.60 [0.36, 1.00] & $-1.94$ ($.052$) & 0.92 [0.62, 1.36] & $-0.42$ ($.674$) & 1.43 [0.86, 2.36] & $1.39$ ($.165$) \\
Reward & 0.39 [0.23, 0.66] & $-3.53$ ($<.001$) & 0.69 [0.46, 1.02] & $-1.88$ ($.060$) & 0.82 [0.50, 1.35] & $-0.78$ ($.433$) \\
\bottomrule
\end{tabular}
\end{table*}

\clearpage

\subsection{Perceived mental effort}
\label{app:expeffort}

We conducted separate one-way ANOVAs comparing perceived mental effort across the five conditions during learning and the unaided test. Perceived effort differed by condition during learning ($F(4, 699) = 2.80$, $p = .025$) and the unaided test ($F(4, 699) = 3.30$, $p = .011$). Tukey-adjusted pairwise comparisons showed lower effort under metacognitive feedback than under the reward in both phases ($\Delta = -0.60$ and $-0.57$, respectively; both $p = .010$). During the final test, effort was also lower in the no-AI control than in the effort-based reward condition ($\Delta = -0.53$, $p = .023$); all other pairwise comparisons were nonsignificant (Table~\ref{tab:effort-contrasts}).

\begin{table*}[!ht]
\centering
\small
\caption{Pairwise differences in perceived mental effort by condition.
Differences are first minus second condition on the 1--9 effort scale.
Confidence intervals and two-sided $p$-values are Tukey-adjusted within each phase.
Omnibus condition effects: learning, $F(4, 699) = 2.80$, $p = .025$; final test, $F(4, 699) = 3.30$, $p = .011$.}
\label{tab:effort-contrasts}
\begin{tabular}{lrcrrcr}
\toprule
& \multicolumn{3}{c}{Learning phase} & \multicolumn{3}{c}{Final test} \\
\cmidrule(lr){2-4} \cmidrule(lr){5-7}
Contrast & $\Delta$ & 95\% CI & $p$ & $\Delta$ & 95\% CI & $p$ \\
\midrule
No-AI control $-$ AI-only                 & $0.08$  & $[-0.42, 0.59]$  & .992 & $-0.20$ & $[-0.68, 0.28]$  & .792 \\
No-AI control $-$ Metacognitive feedback  & $0.39$  & $[-0.11, 0.89]$  & .197 & $0.05$  & $[-0.43, 0.52]$  & .999 \\
No-AI control $-$ Effort-based reward                  & $-0.21$ & $[-0.71, 0.29]$  & .785 & $-0.53$ & $[-1.00, -0.05]$ & .023 \\
No-AI control $-$ Combined                & $0.12$  & $[-0.38, 0.62]$  & .967 & $-0.13$ & $[-0.61, 0.35]$  & .947 \\
AI-only $-$ Metacognitive feedback        & $0.31$  & $[-0.20, 0.82]$  & .448 & $0.25$  & $[-0.24, 0.73]$  & .631 \\
AI-only $-$ Effort-based reward                        & $-0.29$ & $[-0.81, 0.22]$  & .521 & $-0.33$ & $[-0.81, 0.16]$  & .350 \\
AI-only $-$ Combined                      & $0.04$  & $[-0.47, 0.55]$  & $>$ .999 & $0.07$ & $[-0.42, 0.55]$ & .995 \\
Metacognitive feedback $-$ Effort-based reward         & $-0.60$ & $[-1.11, -0.10]$ & .010 & $-0.57$ & $[-1.05, -0.09]$ & .010 \\
Metacognitive feedback $-$ Combined       & $-0.27$ & $[-0.78, 0.23]$  & .571 & $-0.18$ & $[-0.66, 0.30]$  & .851 \\
Effort-based reward $-$ Combined                       & $0.33$  & $[-0.18, 0.84]$  & .394 & $0.40$  & $[-0.09, 0.88]$  & .168 \\
\bottomrule
\end{tabular}
\end{table*}

\clearpage

\subsection{Metacognitive accuracy and metacognitive sensitivity}
\label{app:expmetacognition}

We assessed \textbf{metacognitive accuracy} as signed prediction error of correctly solved test items (actual minus predicted final-test score, range $-6$ to 6), with positive values indicating underestimation, and compared conditions using a one-way ANOVA. Mean prediction errors were positive across conditions, but the condition effect was nonsignificant ($F(4, 699) = 1.44$, $p = .219$; Table~\ref{tab:metacognitive-accuracy}). 

We assessed \textbf{metacognitive sensitivity} using participant-level AUCs relating confidence to response correctness, excluding participants with only correct or only incorrect responses because AUC is undefined when only one response category is present. Mean AUCs exceeded chance in all conditions (all unadjusted $p < .001$), but did not differ significantly between conditions ($F(4, 396) = 0.60$, $p = .662$; Table~\ref{tab:metacognitive-sensitivity}).

\begin{table}[!ht]
  \centering
  \caption{Metacognitive accuracy, operationalized as signed
    prediction error, by condition ($N = 704$).
    Entries are raw means and standard deviations.
    The omnibus condition effect was not significant,
    $F(4, 699) = 1.44$, $p = .219$.}
  \label{tab:metacognitive-accuracy}
  \begin{tabular}{lrrr}
    \toprule
    Condition & $n$ & $M$ & $SD$ \\
    \midrule
    No-AI control          & 146 & 0.93 & 1.63 \\
    AI-only                & 137 & 0.53 & 1.60 \\
    Metacognitive feedback & 144 & 0.60 & 1.71 \\
    Effort-based reward                 & 138 & 0.68 & 1.67 \\
    Combined               & 139 & 0.55 & 1.56 \\
    \bottomrule
  \end{tabular}
\end{table}

\begin{table*}[!ht]
  \centering
  \caption{Participant-level metacognitive sensitivity (AUC)
    by condition. Entries are raw means and standard deviations.
    Participants with only correct or only incorrect test responses
    were excluded because AUC requires both response types.
    Two-sided one-sample $t$-tests compare each condition's mean
    AUC with chance ($0.5$); reported $p$-values are unadjusted.
    The omnibus condition effect was not significant,
    $F(4, 396) = 0.60$, $p = .662$.}
  \label{tab:metacognitive-sensitivity}
  \begin{tabular}{lrrrrrrc}
    \toprule
    Condition & Included & Excluded & $M$ & $SD$
      & $t$ & $df$ & $p$ \\
    \midrule
    No-AI control          & 82 & 64 & 0.75 & 0.28
      & 8.03 & 81 & $< .001$ \\
    AI-only                & 74 & 63 & 0.70 & 0.30
      & 5.87 & 73 & $< .001$ \\
    Metacognitive feedback & 80 & 64 & 0.75 & 0.27
      & 8.11 & 79 & $< .001$ \\
    Effort-based reward                 & 90 & 48 & 0.76 & 0.27
      & 9.16 & 89 & $< .001$ \\
    Combined               & 75 & 64 & 0.72 & 0.29
      & 6.41 & 74 & $< .001$ \\
    \bottomrule
  \end{tabular}
\end{table*}

\clearpage

\subsection{Moderators of metacognitive feedback}
\label{app:traits}

We explored whether (1)~perceived confidence, (2)~trust in AI, and (3)~need for cognition moderated intervention effects on answer offloading and unaided performance using separate item-level mixed-effects logistic regressions for each trait and outcome. Models included the metacognitive feedback$\times$effort-based reward interaction, the mean-centered moderating variable and its interactions with both interventions, item fixed effects, and a participant random intercept. None of the interactions for the potential moderators was significant for either metacognitive feedback (all $p \geq .336$) or effort-based reward (all $p \geq .137$), providing no evidence that either intervention effect varied with these traits (Table~\ref{tab:trait-moderation}).

\begin{table*}[!ht]
\centering
\small
\caption{Moderators of the intervention effects.
Ratios of odds ratios are reported with 95\% confidence intervals and two-sided $p$-values.
Potential moderators were mean-centered. Each block reports one model per outcome. Models included the main effects of feedback and reward, their interaction, the main effect of the moderator and its interactions with both interventions, item fixed effects, and a participant random intercept. Only the moderator$\times$intervention interactions are shown. ROR: ratio of odds ratios. Answer offloading: $N_{\text{obs}} = 5{,}022$; unaided performance: $N_{\text{obs}} = 3{,}348$;
both $N_{\text{participants}} = 558$.}
\label{tab:trait-moderation}
\begin{tabular}{lccrccr}
\toprule
& \multicolumn{3}{c}{Answer offloading} & \multicolumn{3}{c}{Unaided performance} \\
\cmidrule(lr){2-4} \cmidrule(lr){5-7}
Predictor & $\mathit{ROR}$ [95\% CI] & $z$ & $p$ & $\mathit{ROR}$ [95\% CI] & $z$ & $p$ \\
\midrule
\multicolumn{7}{l}{\textit{Perceived confidence}} \\
\quad $\times$ Metacognitive feedback & 1.28 [0.78, 2.10] & $0.96$ & $.336$ & 0.93 [0.68, 1.26] & $-0.50$ & $.619$ \\
\quad $\times$ Effort-based reward & 1.29 [0.78, 2.12] & $0.99$ & $.323$ & 0.84 [0.62, 1.14] & $-1.13$ & $.258$ \\
\addlinespace
\multicolumn{7}{l}{\textit{Trust in AI}} \\
\quad $\times$ Metacognitive feedback & 1.30 [0.61, 2.78] & $0.68$ & $.500$ & 0.84 [0.54, 1.30] & $-0.79$ & $.431$ \\
\quad $\times$ Effort-based reward & 0.99 [0.46, 2.12] & $-0.02$ & $.986$ & 1.19 [0.76, 1.84] & $0.76$ & $.449$ \\
\addlinespace
\multicolumn{7}{l}{\textit{Need for cognition}} \\
\quad $\times$ Metacognitive feedback & 0.65 [0.26, 1.62] & $-0.93$ & $.355$ & 0.80 [0.47, 1.36] & $-0.83$ & $.404$ \\
\quad $\times$ Effort-based reward & 0.51 [0.20, 1.27] & $-1.45$ & $.148$ & 1.49 [0.88, 2.53] & $1.49$ & $.137$ \\
\bottomrule
\end{tabular}
\smallskip
\begin{minipage}{0.9\textwidth}
\footnotesize \textit{Note.} RORs indicate changes in the trait–outcome association associated with each intervention.
\end{minipage}
\end{table*}

\end{document}